\documentclass[aps,showpacs,nofootinbib,longbibliography,superscriptaddress,notitlepage]{revtex4-1}

\usepackage{graphicx}
\usepackage{amsmath}
\usepackage{amsfonts}
\usepackage{amssymb}
\usepackage{slashed}
\usepackage{MnSymbol}
\usepackage{mathbbol}
\usepackage{color}

\usepackage{tikz}
\usetikzlibrary{trees}
\usetikzlibrary{decorations.pathmorphing}
\usetikzlibrary{decorations.markings}
\usetikzlibrary{shapes.misc}
\usetikzlibrary{matrix}
\usetikzlibrary{decorations.pathreplacing}

\usepackage[hyperfootnotes=true]{hyperref}

\newcommand{\refc}[1]{(\ref{#1})}

\newcommand{\dd}{\textmd{d}}
\newcommand{\be}{\begin{equation}}
\newcommand{\ee}{\end{equation}}
\newcommand{\expv}[1]{\left \langle #1 \right \rangle}

\newcommand{\ve}{\mathbf}
\newcommand{\U}{\text{U}}
\newcommand{\SU}{\text{SU}}
\newcommand{\mpc}{\widetilde{m}}
\newcommand{\mb}{m}

\newcommand{\p}{\psi}
\newcommand{\bp}{\bar{\psi}}
\newcommand{\debp}{\delta\bar{\psi}}
\newcommand{\dep}{\delta\psi}
\newcommand{\um}{U^G_\mu}
\newcommand{\dum}{(U^G)^\dagger_\mu}
\newcommand{\uam}{u_\mu}

\newcommand{\Duam}{(\delta u)_\mu}
\newcommand{\Dduam}{(\delta u)^\dagger_\mu}

\newcommand{\csw}{c_{\rm SW}}
\newcommand{\cem}{c_{\rm em}}

\newcommand{\vda}{\alpha_V^j(x)}
\newcommand{\ada}{\alpha_A^j(x)}

\newcommand{\gm}{\gamma_\mu}
\newcommand{\gf}{\gamma_5}
\newcommand{\hmu}{\hat{\mu}}

\newcommand{\com}[2]{\big[ #1 \,,\, #2 \big]}
\newcommand{\acom}[2]{\big\{ #1 \,,\, #2 \big\}}
\newcommand{\etjk}{\epsilon_{3jk}}

\newcommand{\sz}{s_z}
\newcommand{\pz}{p_z}
\newcommand{\Ord}{\mathcal{O}}

\newcommand{\Regensburg}{Institute for Theoretical Physics, Universit\"at Regensburg, D-93040 Regensburg, Germany.}
\newcommand{\Tata}{Tata Institute of Fundamental Research, Homi Bhabha Road, Mumbai 400005, India.}
\newcommand{\Frankfurt}{Institute for Theoretical Physics, Goethe University, Max-von-Laue-Strasse 1, 60438 Frankfurt am Main, Germany}

\begin{document}

\title{Meson masses in electromagnetic fields with Wilson fermions}

\author{G.~S.~Bali}
\email{gunnar.bali@ur.de}
\affiliation{\Regensburg}
\affiliation{\Tata}

\author{B.~B.~Brandt}
\email{brandt@th.physik.uni-frankfurt.de}
\affiliation{\Frankfurt}

\author{G.~Endr\H{o}di}
\email{endrodi@th.physik.uni-frankfurt.de}
\affiliation{\Frankfurt}

\author{B.~Gl\"a\ss{}le}
\email{benjamin.glaessle@ur.de}
\affiliation{\Regensburg}

\begin{abstract}
We determine the light meson spectrum in QCD in the presence of background
magnetic fields using quenched Wilson fermions. Our continuum extrapolated
results indicate a monotonous reduction of the
connected neutral pion mass as the magnetic field grows. The
vector meson mass is found to remain nonzero, a finding relevant for
the conjectured $\rho$-meson condensation at strong magnetic fields. The continuum
extrapolation was facilitated by adding a novel magnetic field-dependent improvement term
to the additive quark mass renormalization. Without this term, sizable lattice
artifacts that would deceptively indicate an unphysical rise of the connected
neutral pion mass for strong magnetic fields are present. We also investigate the impact of
these lattice artifacts on further observables like 
magnetic polarizabilities and discuss the magnetic field-induced 
mixing between $\rho$-mesons and pions. We also derive Ward-Takashi identities for
QCD+QED both in the continuum formulation and for (order $a$-improved) Wilson fermions.
\end{abstract}


\maketitle

\section{Introduction}

Background magnetic fields have a decisive impact on the physics of quarks 
and gluons and offer a wide range of applications. 
Strong magnetic fields appear in noncentral heavy-ion
collisions~\cite{Kharzeev:2004ey,Kharzeev:2007jp}, inside
magnetars~\cite{Duncan:1992hi}, and might have been generated during 
the evolution of the early Universe~\cite{Vachaspati:1991nm}. An important characteristic 
of the magnetized QCD medium -- relevant for each of the above
settings -- is the effect of the magnetic field on the spectrum of the theory.
It has been speculated~\cite{Bonati:2015dka,Fukushima:2016vix} that
long-lived magnetic fields might affect hadronization in 
heavy-ion collisions. In strongly magnetized neutron stars
the magnetic properties of QCD matter, predominantly those of the light hadrons,
may have a significant impact on the mass-radius relation~\cite{Broderick:2000pe}.

The direct influence of the external field on the phase
structure of QCD has received considerable attention recently; see for instance the
review~\cite{Andersen:2014xxa}. In particular, lattice
simulations have revealed that the transition temperature of the chiral
restoration/deconfinement crossover monotonously decreases as the magnetic field
grows~\cite{Bali:2011qj,Endrodi:2015oba}; see
also Refs.~\cite{Bornyakov:2013eya,Ilgenfritz:2013ara}. Thinking in terms of a hadronic
model, in which the temperature needed to resolve the quark structure of a hadron
is of the order of the hadron mass, it is plausible that the magnetic
field dependence of the masses of the lightest hadrons do play a relevant role
for the structure of the phase diagram.
Specifically, pions are expected to be dominant in this respect. It is well
known from chiral perturbation theory~\cite{Andersen:2012dz} that charged and 
neutral pions respond oppositely to the magnetic field $B=|\ve{B}|$. 
While lattice results employing staggered~\cite{Bali:2011qj} and 
quenched unimproved Wilson~\cite{Hidaka:2012mz} quarks agree that the charged pion 
mass $m_{\pi^{\pm}}$ is increased by the presence of a magnetic field, there is a 
discrepancy in the literature about $m_{\pi^0}(B)$ 
for intermediate 
magnetic fields. It monotonously decreases
for quenched overlap quarks~\cite{Luschevskaya:2014lga}, while for quenched
unimproved Wilson fermions, a
turning point and subsequent increase was reported in
Ref.~\cite{Hidaka:2012mz}.\footnote{In fact, the above mentioned lattice results 
are based on the connected contribution to the pion correlator, which corresponds 
to a hypothetic meson with exclusively $\bar u u$ or $\bar d d$ flavor content. 
We get back to this point in Sec.~\ref{sec:mixing} below.} 
In light of the above discussion, the two results hint at
opposite implications for the QCD phase diagram for strong magnetic fields.

Clarifying this question may also be relevant for yet another aspect of
the phase structure of QCD
with $B>0$: It has been conjectured that strong magnetic fields could 
reduce the vector meson mass to zero and
lead to the condensation of $\rho$-mesons~\cite{Schramm:1991ex}. They may also induce a 
transition to a superconducting
phase~\cite{Chernodub:2011mc}.\footnote{Note that this
is the simplest scenario for the superconductivity of the QCD
vacuum when exposed to external magnetic fields. 
In fact, when coupled to the photon field that exhibits gauge fluctuations, the 
would-be massless $\rho$ mode is absorbed by the photon field in accordance with the 
Higgs mechanism~\cite{Chernodub:2013uja} so that no massless mode remains. 
For constant background magnetic fields -- like the one 
we are working with here -- this mechanism is absent. 
Also note that the proposed superconducting phase was suggested to exhibit several unusual features
like anisotropy and inhomogeneity~\cite{Chernodub:2010qx,Chernodub:2011mc}, and 
we do not attempt to capture these subtle characteristics here, which would require 
dynamical QCD+QED simulations and is much more involved.}
On the one hand, QCD inequalities can be used to show that massless 
$\rho$-mesons are only allowed if the (connected) neutral pion mass
vanishes as well. The existence of a turning point in $m_{\pi^0}(B)$,
above which it grows with increasing $B$, would thus speak against this scenario.
On the other hand, a study using quenched overlap quarks in two-color QCD,
based on spatial vector 
correlators, has reported a critical magnetic field of around $0.9 \textmd{ GeV}^2$, beyond
which $\rho$-condensation was found to occur~\cite{Braguta:2011hq}.

Resolving the above discrepancy regarding $m_{\pi^0}(B)$ is thus clearly of high importance. 
In Ref.~\cite{Bali:2015vua}, we published preliminary results that explain the 
origin of the difference between the quenched overlap and Wilson results.
We have found that Wilson quarks 
are subject to a $B$-dependent additive mass renormalization at finite lattice 
spacings. Although this effect disappears in the continuum limit, 
it has a drastic impact on the meson masses 
evaluated at nonzero lattice spacings. We also sketched a method to
subtract these artifacts by determining a magnetic field-dependent 
line of constant physics for the Wilson bare mass parameter.

In this paper, we expand on the ideas introduced in Ref.~\cite{Bali:2015vua} 
and discuss the improvement of scaling toward the continuum limit 
via the $B$-dependent additive mass renormalization in more detail.
Performing this improvement reveals that the neutral pion mass monotonously
decreases as $B$ grows and, thus, resolves the existing disagreement between
the quenched Wilson and overlap formulations. We also determine the $\rho$-meson mass
and find that it remains nonzero in the whole range of magnetic fields studied,
$0<eB<4\textmd{ GeV}^2$.
For comparison, we also measure the connected neutral pion 
mass using dynamical staggered quarks
on existing configurations from the study of
Ref.~\cite{Bali:2011qj}.\footnote{Note,
that the use of staggered fermions for the spectroscopy of vector mesons at
finite $B$ is complicated, since the magnetic field affects the group theoretical
construction of vector meson interpolators. Furthermore, the use of the staggered
formulation complicates the interpretation of higher-lying states due to taste splitting.}

There are further interesting quantities that are 
related to the spectrum, including 
magnetic polarizabilities and magnetic moments. These describe
the leading-order response of hadrons to the
external field and, thus, involve derivatives at zero field; see, e.g., 
Refs.~\cite{Luschevskaya:2012xd,Luschevskaya:2014lga,Beane:2014ora}. 
In contrast to observables at finite values of the field, these can also 
be defined for background electric fields, giving rise to electric polarizabilities.
We will show that for Wilson quarks
both types of polarizabilities suffer from enhanced lattice artifacts that 
can (and should) be eliminated using our improvement program.

We emphasize that our results are obtained in the quenched approximation, just 
as all studies in the literature so far 
concerning the light meson spectrum at finite $B$ (except for 
the staggered results of Ref.~\cite{Bali:2011qj},
which, however, focused on pions only).
The quenching of virtual quark loops 
induces a systematic effect that is hard to estimate. 
At $B=0$ it is known that the results for the quenched and the dynamical spectrum agree 
within about 10\% (see Refs.~\cite{Aoki:1999yr,Aoki:2002uc},
for instance), indicating that the meson masses are dominantly determined by
the properties of valence quarks. This is expected to remain true in the presence
of external fields. For example the mass of the neutral pion, being a pseudo-Goldstone
boson even for $B>0$, depends primarily on the quark masses and not on gluonic properties. 
For further mesons one may expect a moderate quenching effect
as long as the vacuum exhibits a structure similar to that at $B=0$.
Observables related to the bulk properties of the gluonic field (such as the static
quark-antiquark potential considered in Ref.~\cite{Bonati:2016kxj}), however, are expected
to show stronger quenching effects, since the primary effect of the external field
enters only indirectly via the sea quark loops which are not present in the quenched setup.

This paper is organized as follows. We start with a few remarks about the meson spectrum 
in continuum QCD in the presence of background magnetic fields in Sec.~\ref{contiB}. 
This is followed by Sec.~\ref{sec:latticestuff}, where our lattice setup is described and 
our improvement scheme relevant for Wilson fermions is detailed. Section~\ref{sec:results}
contains our results about the spectrum and the analysis of the dependence on quark masses 
and on the lattice spacing. In Sec.~\ref{polar} we consider the impact of our improvement 
scheme for magnetic polarizabilities before we conclude in Sec.~\ref{sec:concl}. 
Three Appendixes contain our results in the free case (Appendix~\ref{app1}), the derivation of Ward-Takahashi 
identities for nonzero electromagnetic fields (Appendix~\ref{app2}) and a perturbative check 
of the $B$-independence of the multiplicative mass renormalization constant in QCD
(Appendix~\ref{app3}).

\section{Meson spectrum in the continuum}
\label{contiB}

\subsection[Meson and quark masses for $B>0$]{Meson and quark masses for \boldmath$B>0$}

We begin with a few general comments about the magnetic field dependence 
of the pion masses in the continuum. For small magnetic fields (and at zero temperature)
we may approximate the charged pion as a pointlike free scalar particle
with electric charge $\pm e$. In that case,
the quantum mechanically allowed energies are the Landau levels,
\be
E_n^2 = m_\pi^2 + (2n+1)|eB| + \pz^2 , \quad\quad n\in\mathbb{Z}_0^+\,,
\label{eq:Escalar}
\ee
where $m_\pi$ is the $B=0$ pion mass and we have assumed $\ve{B}=B\,\ve{e}_z$.
Thus, the magnetic field-dependent mass (the lowest energy with $n=\pz=0$) reads
\be
m_{\pi^\pm}(B) = \sqrt{m_\pi^2 + |eB|}\,.
\label{eq:LLb}
\ee
The same argument for the neutral pion gives a magnetic field-independent mass,
$m_{\pi^0}(B)=m_\pi$. Thus, the neutral pion remains massless for nonzero 
magnetic fields if $m_\pi=0$, while the charged pions become massive for $B\neq0$. 

It turns out that the conclusion of increasing charged pion masses and an almost constant
neutral pion mass remains valid if the interaction between 
pions is taken into account. This can be done consistently within chiral 
perturbation theory for nonzero magnetic
fields~\cite{Shushpanov:1997sf,Agasian:2001ym,Andersen:2012dz,Andersen:2012zc}.
Indeed, the neutral pion remains a Goldstone boson even for $B\neq0$. 
In addition, away from the chiral limit, the Gell-Mann-Oakes-Renner relation
\be
m_f = \frac{m_{\pi_0}^2 F_{\pi^0}^2}{2\bar\psi\psi} + \Ord(m_{\pi_0}^4)
\label{eq:gmor}
\ee
still applies and is independent of
$B$~\cite{Shushpanov:1997sf,Agasian:2001ym}. 
Here $F_{\pi^0}$ is the decay constant of the neutral pion, $\bar\psi\psi$
is the (in our convention, positive) chiral condensate and $m_f$ denotes the (degenerate) up and down 
quark masses. $F_{\pi_0}$ and $\bar\psi\psi$ above are defined in the chiral limit, $m_{\pi_0}\to0$.
Equation~(\ref{eq:gmor}) is consistent with the notion that the quark mass does not depend on the magnetic field.

The independence of the quark masses on the magnetic field also becomes obvious
in a completely different limit: at very high temperatures.
Here, quarks become quasifree due to asymptotic freedom. 
Then the quantum mechanical energy levels are the Landau levels for fermions
with charge $q$,
\be
E_{f;n}^2 = m_f^2 + 2n|q_fB| + \pz^2, \quad\quad n\in\mathbb{Z}_0^+\,,
\label{eq:Efermion}
\ee
so that the lowest energy ($n=\pz=0$) is indeed magnetic
fieldindependent.\footnote{Clearly, if QCD interactions are present, the
levels will mix. Nevertheless, it turns out that the lowest Landau level can
still be defined unambiguously and remains approximately
$B$ independent~\cite{Bruckmann:2016bcl,Bruckmann:2017pft}.}
The free case is of interest for yet another reason: In the strong magnetic field 
limit $B\to\infty$, quarks and gluons decouple and pure fermionic 
observables approach their free-case values due to asymptotic freedom
(see, for example, Refs.~\cite{Endrodi:2015oba,Miransky:2015ava}). It is therefore
instructive to compare our QCD results at large magnetic fields to the free case;
see below.

In QCD, the quark mass is also subject to a multiplicative renormalization 
of the form $m_f^r=Z_m m_f$. Since $Z_m$ is related to the ultraviolet behavior 
of the theory, while the magnetic field is a physical infrared parameter,
one expects that $Z_m$ is also independent of $B$. We demonstrate this by an
explicit calculation using perturbation theory in Appendix~\ref{app3}.

In the following, we will also investigate the effect of the magnetic field 
on $\rho$-mesons. 
Unlike for pions, for vector mesons, the magnetic field interacts 
directly with the spin of the particle.
In the free case, the energies of a pointlike vector meson (with 
a gyromagnetic ratio $g=2$ and charge $q=\pm e$) are given
by~\cite{Chernodub:2010qx}
\be
\label{eq:enes-spin}
E^2_n = m_\rho^2 + (2n+1)|eB|-g\sz qB + \pz^2 \,,
\ee
where $\sz=0,\pm1$ is the projection of the spin on the magnetic field axis.
For $\sz=1$ one may expect the mass of
the associated $\rho$-meson to vanish when $eB=m_\rho^2$.
This has lead to the proposal that the $\rho$-mesons could
condense~\cite{Schramm:1991ex} and initiated the investigation of the possibility of a 
superconducting vacuum for strong magnetic fields~\cite{Chernodub:2011mc}, as mentioned in the
Introduction. However, QCD inequalities imply~\cite{Hidaka:2012mz} that the
correlation function in the vector channel
is bounded from above by the (connected)
neutral correlation function of the pion, so that
\be
\label{eq:ineq-bound}
m_\rho\geq (m_{\pi^u}+m_{\pi^d})/2\,.
\ee
Thus, a vanishing $\rho$-meson mass
is only possible when $m_{\pi^u/\pi^d}=0$.

Note that the $\rho$-meson is a resonance in nature, rather than a stable particle,
so the extraction of its resonant mass from lattice simulations demands a careful
finite size analysis. In our study, however, the $\rho$-meson can be considered to be
a stable particle, since we are working in the quenched approximation.

\subsection[Mixing of meson states for $B>0$]{Mixing of meson states for \boldmath$B>0$}
\label{sec:mixing}

When we consider the neutral pion at $B=0$, where the $\SU_V(2)$ symmetry
is intact, the associated state is a mixture of neutral pions with
$\bar uu$ and $\bar dd$ flavour contents,
\be
\label{eq:pion-mixB0}
\big| \pi^{0} \big> = \frac{1}{\sqrt{2}} \Big( \big| \pi^{u} \big> - \big|
\pi^{d} \big> \Big) \,.
\ee
At $B>0$, however, $\SU_V(2)$ is broken explicitly by the different
quark electric charges, $q_d=-q_u/2=-e/3$, where $e>0$ is the elementary charge. 
In this case, the
neutral pion will be given by the more general relation
\be
\label{eq:pion-mix}
\big| \pi^0 \big> = \alpha(B) \big| \pi^{u}
\big> - \beta(B) \big| \pi^{d} \big> \quad \text{with} \quad
\alpha^2(B)+\beta^2(B)=1 \,,
\ee
where $\alpha(B)\to1/\sqrt{2}$ and $\beta(B)\to1/\sqrt{2}$ for $B\to0$. 
Thus, determining quantities related to the neutral pion 
in principle involves the computation of the coefficients
$\alpha$ and $\beta$ for each value of the magnetic field. Note that for $B>0$, the
associated correlation matrix
contains disconnected diagrams (these terms cancel at $B=0$ due to isospin symmetry).
In the present study, we neglect disconnected diagrams and the associated mixing 
and work with the individual $\pi^u$ and $\pi^d$-states instead.
This approximation is expected to become valid for strong magnetic fields, 
where the free case is approached (see the discussion above) and, thus, 
disconnected diagrams become negligible.
Note that the so-defined connected pion states $\pi^u$ and $\pi^d$ are 
still useful as, for example, they enter in QCD inequalities~\cite{Hidaka:2012mz}.
We remark furthermore that neutral $\rho$-mesons are affected by the above 
issue in the same way. 

The presence of a finite external magnetic field enables the mixing between the
pion and the $\rho$-meson with $\sz=0$,
\be
\label{eq:rho-decay}
B\neq0: \qquad \rho^{\pm,0}_{\sz=0} \equiv \rho^{\pm,0}_{0} \quad
\longleftrightarrow \quad \pi^{\pm,0} \,.
\ee
Throughout the paper, superscripts of meson states denote electric charge, and
subscripts denote the spin.
For $\sz=\pm1$, the above mixing is forbidden due to the conservation of angular
momentum. Consequently, pion states contribute to the spectral
representation of correlation functions associated with $\sz=0$ $\rho$-mesons,
and, in turn, these $\rho$-meson states also contribute to the spectral
representation of pion correlation functions, so both correlation
functions show the same leading-order exponential falloff. In fact, once this
mixing is enabled, the physical mass eigenstates are mixtures of would-be $\pi$
and $\rho$-states which can be written as
\be
\label{eq:pi-rho-mix}
\big| (\pi')^{\pm,0} \big> = \cos(\theta) \big| \pi^{\pm,0} \big> +
\sin(\theta) \big| \rho^{\pm,0}_{0} \big> \quad \text{and}
\quad \big| (\rho')^{\pm,0}_{0} \big> = -\sin(\theta) \big| \pi^{\pm,0} \big>
+ \cos(\theta) \big| \rho^{\pm,0}_{0} \big> \,,
\ee
where $\theta$ is the mixing angle and we denote the state with smaller energy
by $\pi'$. In principle, the corresponding energies can be extracted from a
correlation matrix using the operators for $\pi^{\pm,0}$ and $\rho^{\pm,0}_{0}$,
while $\theta$ will depend on the interpolators in use. On top of these two
(and higher excited single-particle) states, 
there are also states containing more pions (and radial excitations) 
that may be involved in the mixing. However, their contribution is 
negligible if $m_{\rho'}<E_1$ is satisfied, where $E_1$ is the energy of the
lowest-lying multiparticle excitation.

\section{Magnetic fields and quark masses with Wilson fermions on the lattice}
\label{sec:latticestuff}

\subsection{Lattice setup}

We consider two flavors of (quenched) unimproved Wilson fermions on
the lattice with the fermion action
\be
\label{eq:Wilson-act}
S_F = a^4 \sum_x \Big[ \bar{\psi}(x) \Big(M^0+\frac{4r}{a}\Big) \psi(x) -
\frac{1}{2a} \sum_{\mu=0}^{4} \Big( \bar{\psi}(x) (r-\gamma_\mu) U_\mu(x)
\psi(x+a\hat{\mu}) + \bar{\psi}(x+a\hat{\mu}) (r+\gamma_\mu) U_\mu^\dagger(x)
\psi(x) \Big) \Big] .
\ee
Here $a$ denotes the lattice spacing, $\psi=(u,d)^\top$ is the Dirac field
for flavor $u$ and $d$, and $M^0$ is the
matrix of bare quark masses,
\be
\label{eq:mass-mat}
M^0 = \frac{1}{2} (m^0_u+m^0_d) \mathbf{1} + \frac{1}{2} (m^0_u-m^0_d) \tau^3 \,,
\ee
where $\mathbf{1}$ is the unit matrix in flavor space and $\tau^3$ is the third
Pauli matrix. $\gamma_\mu$ denote the Euclidean Dirac matrices.
$r$ is the coefficient of the Wilson term which we will set to
$r=1$ in the simulations.
The quarks couple to gluonic and electromagnetic fields via the
link variables $U_\mu(x)$, containing the gluonic links $U_\mu^{\rm
G}(x)\in\SU(N)$ and the electromagnetic link variables $u_\mu(x)$,
\be
\label{eq:u3-links}
 U_\mu(x) = U_\mu^{\rm G}(x) u_\mu(x) \, \in \, \U(N)\times\SU_f(2) \quad
\text{with} \quad u_\mu(x) = \exp\Big[ i e a
\Big(\frac{\mathbf{1}}{6}+\frac{\tau^3}{2}\Big) A_\mu(x)\Big] .
\ee
The latter are matrices in flavor space since the electromagnetic charges
are different. We are mostly interested in QCD in
the presence of an external magnetic field $\ve{B}=B \, \ve{e}_z$.
Such a magnetic field can be
generated by a vector potential of the form
\be
\label{eq:vector-pot}
\begin{array}{c}
A_x(x)=-\eta B y - (1-\eta) B y L_x \delta(x-L_x) \,, \quad
A_y(x)=(1-\eta) B x + \eta B x L_y \delta(y-L_y) \\ \text{and} \quad
A_t(x)=A_z(x) = 0 \,,
\end{array}
\ee
where the parameter $\eta$ cancels in
physical observables and thus can be chosen at will. In the following
results are obtained using the ``symmetric gauge,'' $\eta=1/2$, but we have
explicitly checked that the results agree when we set $\eta=1$. $L_\mu=N_\mu a$
denotes the spatial extent in the $\mu$-direction and $N_\mu$ is the
number of lattice points. The boundary $\delta$-terms ensure periodic boundary
conditions for the gauge potential and the constancy of the magnetic field in
the $x,y$-plane~\cite{AlHashimi:2008hr}. Note, that in a finite volume, the
magnetic field is quantized according to
\be
\label{eq:fquant}
q_dB \, N_x N_y = 2 \pi N_b , \quad \text{with} \quad N_b\in\mathbb{Z} \quad
\text{and} \quad 0\leq N_b < N_x N_y \,,
\ee
where the smallest quark charge, $q_d=-e/3$ appears.

\subsection{Quark mass renormalization}

The term proportional to $r$ in Eq.~(\ref{eq:Wilson-act})
is the well-known Wilson term, which is introduced
to remove the doublers from the theory. This term
breaks chiral symmetry explicitly so that the quark mass is subject to additive
renormalization, besides the multiplicative renormalization that is already necessary 
in the continuum. 
From the arguments of Sec.~\ref{contiB}, we expect the
multiplicative renormalization factor $Z_m$ to be independent of the external field 
-- or, in other words,
a $B$-independent renormalization scheme can be chosen (see Appendix~\ref{app3}).
Thus, at finite lattice spacing, the renormalized quark mass of
flavor $f$ is given by
\footnote{Note that a similar equation holds for the case of the coupling
to a dynamical electromagnetic field (e.g. Ref.~\cite{Borsanyi:2014jba}),
\begin{displaymath}
 m_f^r=Z_m(a,e) \big[m^0_f-m_{c;f}(a,e)\big] \,,
\end{displaymath}
where both renormalization factors depend on the bare electromagnetic coupling
$e$.} 
\be
m_f^r=Z_m(a) m_f = Z_m(a) \big[m^0_f-m_{c;f}(a,B)\big] \,,
\label{eq:massreno}
\ee
where
\be
\label{eq:mc-Bdep}
am_{c;f}(a,B) = am_{c}(a,0) + d_m a^2 |q_fB| + \Ord\big([a^2 |q_fB|]^2\big)\,, 
\ee
and $d_m$ is the coefficient of the leading-order $B$-dependent term.
Note that, although $d_m$ corrects
for a lattice artifact, it is not a Symanzik improvement coefficient.
However, it can have a nontrivial value since our theory is not
$\Ord(a^2)$ [and not even $\Ord(a)$] improved.
In Appendix~\ref{app1} we give an
intuitive explanation of why magnetic fields shift the
quark mass, i.e., why $m_{c;f}$ becomes $B$ dependent. 
The argument boils down to the fact that the Wilson term coincides with the action of a
scalar particle -- multiplied by the lattice spacing so that it vanishes for
$a\to0$. In the presence of the magnetic field, the squared mass of this scalar
particle is shifted by $B$ according to Eq.~(\ref{eq:LLb}). Eventually this
induces the $B$-dependence in the additive mass renormalization. This can be
checked explicitly in the free case (where $d_m=-1/2$), for which we show results in
Appendix~\ref{app1}.

It is worth stressing that the presence of the additive quark mass renormalization
is a lattice artifact and $m_{c;f}(a,B)\to0$ for $a\to0$ for any magnetic field. For Wilson fermions
it is customary to introduce the hopping parameter $\kappa_f$ and its critical value
$\kappa_{c;f}$,
\be
\label{eq:kappas}
\kappa_f = (2am^0_f+8)^{-1} \quad \text{and} \quad \kappa_{c;f}(a,B) =
(2am_{c;f}(a,B)+8)^{-1} \,,
\ee
so that the $B$-dependent quark mass renormalization can be translated to the $B$-dependence 
of $\kappa_{c,f}$,
\be
\label{eq:kc-Bdep}
\frac{1}{\kappa_{c;f}(a,B)} = \frac{1}{\kappa_{c}(a,0)} +2d_m a^2 |q_fB| + \Ord\big([a^2 |q_fB|]^2\big)\,.
\ee

At finite lattice spacing the $B$-dependence of $m_{c;f}$ contaminates physical
$B$-effects. Due to this additional lattice artifact, very
fine lattices are necessary for a reliable continuum extrapolation. A better
way is to get rid of the contamination by keeping the simulation on the line of
constant physics (LCP), characterized by a constant renormalized quark mass $m_f^r$.
This LCP depends on the magnetic field according to Eq.~\refc{eq:massreno},
and will be referred to as LCP(B) from now on.\footnote{We
remark that, for unimproved Wilson fermions, lines of constant physics are,
in general, only defined up to
magnetic field-dependent or -independent $\Ord(a)$ lattice artifacts. Here, we choose
to correct for the $B$-dependent artifacts originating from the additive mass
renormalization, since these turn out to be numerically large.}
Since $Z_m$ is, in our definition, $B$ independent, it suffices to tune the
hopping parameter $\kappa_f=\kappa_f(a,B)$ such that the bare
(multiplicatively unrenormalized) quark mass
\be
\label{eq:bm-def}
\mb_f \equiv \big[m^0_f(a,B)-m_{c;f}(a,B)\big] =
\frac{1}{2a}\bigg[\frac{1}{\kappa(a,B)} - \frac{1}{\kappa_{c;f}(a,B)} \bigg]
\ee
remains constant when varying $B$. We will
discuss the impact of this tuning on lattice artifacts in detail in
Sec.~\ref{compare}, where we compare the results obtained when keeping $\kappa_f$
constant while varying $B$ to those obtained along the LCP(B).
The $B$-dependence of the quark mass is also inherited by further 
quantities that are related to derivatives with respect to the magnetic field.
Particular examples, namely magnetic moments and polarizabilities,
will be discussed in Sec.~\ref{polar}.
 
Before we move on, let us collect some theoretical expectations concerning the
behavior of $m_{c;f}(a,B)$. Parity symmetry ensures that $m_{c;f}(a,B)=m_{c;f}(a,-B)$.
Below we show numerical results in the quenched setup, where charged quarks are only
included in the valence sector. In this case $|q_u/q_d|=2$ leads to
\be
\label{eq:der-rel}
m_{c;u}(a,B)=m_{c;d}(a,2B) \quad \text{or} \quad
\left.\frac{\partial
m_{c;u}(a,B)}{\partial B}\right|_{B=0} = 2
\left.\frac{\partial
m_{c;d}(a,B)}{\partial B}\right|_{B=0} \,.
\ee

\subsection{Determination of the critical mass}

The key ingredient to fixing the LCP(B) is the
$B$-dependence of $\kappa_{c;f}$. To determine it, we can make use of the fact
that the neutral pion mass vanishes in the chiral limit, even at finite $B$ (cf.\
Sec.~\ref{contiB}).
In our quenched setup this chiral extrapolation is complicated by
the appearance of chiral logarithms in quenched chiral perturbation
theory~\cite{Sharpe:1992ft,Bernard:1992mk}
(see also the right panel of Fig.~\ref{fig:awi-vs-kappa} below).
It is more convenient to use the current quark
mass $\mpc_f$ obtained from the axial Ward-Takahashi identity (WI). 
In Appendix~\ref{app2} we derive the
axial and vector WIs in the continuum and for Wilson fermions for $B>0$. While
the charged WIs acquire additional $B$-dependent terms,\footnote{See
also~\cite{Scherer:2002tk,Blum:2007cy,Giusti:2017jof}.} the
neutral WIs are found to keep their $B=0$ form,
\be
\label{eq:PCAC-mass}
a\mpc_f(B) = \frac{\partial_0 \big< (J_A)^{f}_0(x_0) P^{f}(0)
\big>}{2\,\big< P^{f}(x_0) P^{f}(0) \big>} \quad \text{with} \quad x_0\neq0
\,.
\ee
Here, $(J_A)^{f}$ is the (point-split) axial current operator with fermionic fields
of flavors $f$ (cf. Eq.~\refc{eq:psA}), and $P^{f}$ the associated pseudoscalar density.
The isosinglet WI suffers from the well-known axial anomaly, see Eq.~\refc{eq:AWI-conti}, which
involves only gluonic contributions for background magnetic fields, and obtains contributions from
disconnected diagrams, which we neglect. This corresponds to pretending that we have two
different flavors carrying the same electric charge, in analogy to replacing $\pi^0$ by $\pi^u$,
thereby avoiding disconnected diagrams. In the following, we will thus determine $\kappa_{c;f}$
by requiring that the ``neutral'' current quark mass $\mpc_f$, defined via
Eq.~\refc{eq:PCAC-mass}, vanishes at $\kappa_{c;f}(B)$ for each value of $B$.

The correlation functions have been projected to zero momentum in the $z$-direction. 
In addition, we have summed the correlation functions over the
$x,y$-plane at the sink. For $B=0$, this corresponds to projecting to zero
momentum in the $x$- and $y$-directions. On the level of the
electromagnetic vector potential, translational invariance is broken,
and Fourier transformation is not well defined in the $x,y$-plane
(see also Ref.~\cite{Hidaka:2012mz}). 
We found that the summation enhances statistics but does not affect our
results, as we have checked explicitly by comparing the results from summed
correlation functions to those obtained without summation at a fixed position
in the $x,y$-plane. For each value of $a$ and $B$, we measured $\mpc_f$ for
several values of $\kappa$
and determined $\kappa_{c;f}(a,B)$ via a linear extrapolation of the
form [cf.\ Eq.~\refc{eq:bm-def}]
\be
\label{eq:kapc-extra}
\mpc_f(a,B;\kappa) = C_f(a,B) \cdot m_f = \frac{C_f(a,B)}{2a}\,
\bigg[\frac{1}{\kappa}-\frac{1}{\kappa_{c;f}(a,B)}\bigg]
\,.
\ee
Here $C_f$ is a proportionality constant, which relates the two estimates for the bare
quark mass to each other and, thus, may contain ($B$-dependent) lattice artifacts. It
can be written as
\be
\label{eq:cf-Bdep}
C_f(a,B) = Z(a) \Big( 1 + d_Z a^2 |q_fB| + \Ord\big([a^2 |q_fB|]^2\big) \Big) ,
\ee
where we have introduced the $B=0$ proportionality factor (see,
e.g., Ref.~\cite{Fritzsch:2010aw})
\be
\label{eq:Zdef}
Z = \frac{Z_mZ_P}{Z_A}
\ee
and $d_Z$ is the coefficient of the leading-order $B$-dependent lattice artifact
in $C_f$.

We note in passing that ideally one would like to use the charged axial and
vector WIs for the tuning of sums and differences of quark masses. The 
corresponding identities are worked out including electromagnetic interactions
[and also $\Ord(a)$-improvement] in Appendix~\ref{app2}. In this case
no disconnected diagrams contribute. 
We have also looked at the charged WIs but
could not detect plateaus before the signal got lost in the noise. 
A potential reason for this is
that the correlation functions used to construct the charged WIs
decay with a mass larger than the smallest mass in the spectrum
(since $m_{\pi^+}>m_{\pi^0}$)
and thus suffer from the standard signal-to-noise problem. 
We plan to address this and investigate the charged WIs in more detail in the future.

\subsection{Lattice setup and lines of constant physics}
\label{sec:lcpB}

\begin{table}[t]
 \centering
 \begin{tabular}{c|cc|c}
  $\beta$ & Lattice & $r_0/a$ & $\kappa_c(B=0)$ \\
  \hline
  \hline
  5.845 & $36\times12^3$ & 4.020 & 0.16170(4)(46) \\
  6.000 & $48\times16^3$ & 5.368 & 0.15703(2)({ }1) \\
  6.260 & $72\times24^3$ & 8.050 & 0.15234(1)({ }4) \\
  \hline
  \hline
 \end{tabular}
 \caption{\label{tab:runp}Details of the
 quenched ensembles. The results for $\kappa_c(B=0)$ have been obtained
 from the linear extrapolation of the data for $N_b=0$; the first
 uncertainty is statistical, while the second results from the systematic
 uncertainty associated with the extrapolation, estimated by using
 a polynomial of second degree for the extrapolation.}
\end{table}

The simulations are performed in the quenched setup using the Wilson plaquette
action~\cite{Wilson:1974sk} and heat bath and over-relaxation updates with a ratio of 1:4.
The generation of configurations and the measurements of correlation
functions have been carried out employing a modified version of
CHROMA~\cite{Edwards:2004sx}. The run parameters
are collected in Table~\ref{tab:runp}. We also list the value of
the Sommer parameter $r_0/a$ obtained from the interpolation provided
in Ref.~\cite{Necco:2001xg}. To convert to physical units, we use $r_0=0.5$ fm. The
lattice sizes have been tuned to keep the physical volume fixed. This also
allows us to achieve the same physical magnetic field -- complying with the
quantization condition~\refc{eq:fquant} -- for the different lattice spacings.
The correlation functions have been computed using a Wuppertal
smeared~\cite{Gusken:1989qx} source including (three dimensionally) APE
smeared~\cite{Falcioni:1984ei} link variables and a point sink. For the inversions, we have used
the DFL-SAP-GCR solver introduced in Refs.~\cite{Luscher:2003qa,Luscher:2007se}.

\begin{figure}[t]
\begin{minipage}[c]{.47\textwidth}
 \includegraphics[]{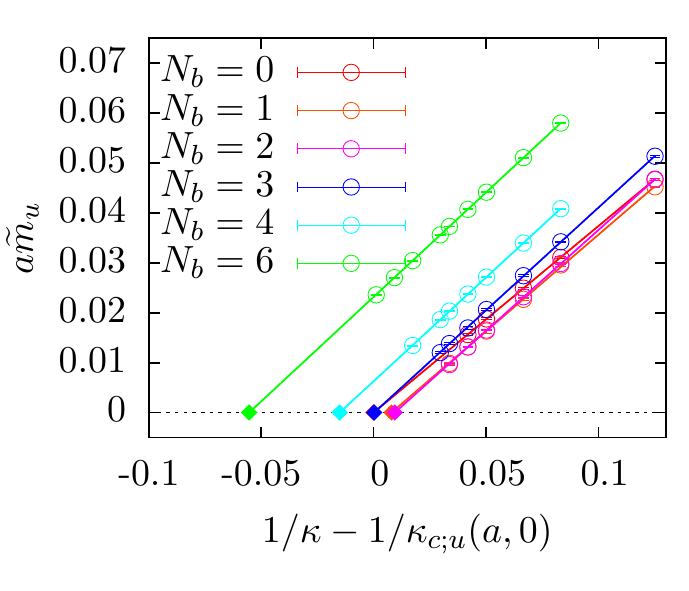}
\end{minipage}
\begin{minipage}[c]{.47\textwidth}
 \includegraphics[]{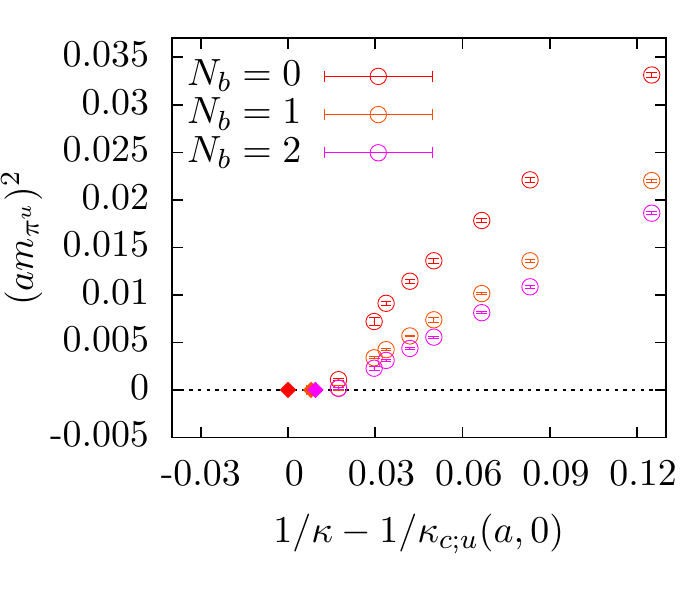}
\end{minipage}
\caption{Results for $a\mpc_u(a,B;\kappa)$ (left) and
$am_{\pi^u}(a,B;\kappa)$ (right) versus $1/\kappa-1/\kappa_{c;u}(a,0)$ for several
values of $N_b$ on the $\beta=6$ ensemble. The colored bands in the left
figure are the results from linear fits and the diamonds on the
$a\mpc_u(a,B;\kappa)=0$ line in both panels are the results for
$1/\kappa_{c;u}(a,B)-1/\kappa_{c;u}(a,0)$. The error bars of the extrapolated
results include the estimate for the systematic uncertainty explained in the
text and are smaller than the symbols.}
\label{fig:awi-vs-kappa}
\end{figure}

To determine $\kappa_{c,f}$, we have measured $\mpc_u$ and $\mpc_d$ for
several values of $\kappa$ with pion masses between 1~GeV and about 400~MeV on
each of these ensembles and for magnetic fields $0<eB<4$~GeV$^2$.
For $B=0$, the above setup satisfies $m_{\pi^{f}} L\gtrsim 3$ for all measurement points.
However, as we will see below, the neutral pion mass decreases significantly with
increasing magnetic field, so that finite size effects can potentially increase
as $B$ grows. To show that this is not the case, we study finite
size effects in Sec.~\ref{fse}.

\begin{figure}[t]
\begin{center}
 \includegraphics[]{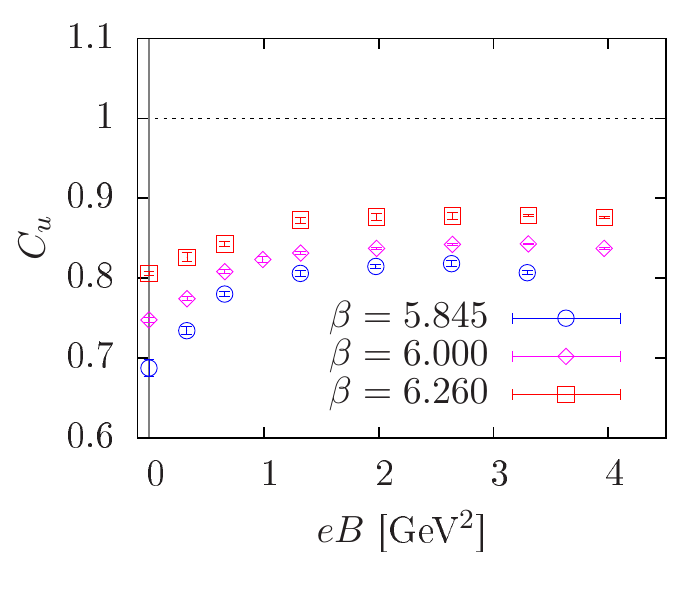}
\end{center}
\caption{Results for $C_u$ versus $|eB|$ for different values of the lattice
spacing. The dashed line indicates the continuum limit value $C_u=1$, and the gray
solid line is the $B=0$ line.}
\label{fig:Z-vs-B}
\end{figure}

We have extracted
$\kappa_{c;f}(a,B)$ using a linear chiral extrapolation of the form given in
Eq.~\refc{eq:kapc-extra}. We show the results for $a\mpc_u(a,B;\kappa)$
together with the chiral extrapolations for several values of $B$ obtained on
the $\beta=6.0$ ensemble in Fig.~\ref{fig:awi-vs-kappa} (left). As can be seen from
the plot, the value of $\kappa_{c;u}(a,B)$ has a strong dependence on $B$, while
the slope $C_u$ is only mildly affected. In fact, the strong change in
$\kappa_c$ forced us to measure $\mpc_u$ for values of $\kappa$ which would have
lead to a negative value for $\mb_u$ at $B=0$. In Fig.~\ref{fig:awi-vs-kappa}
(right), we show the results for the mass of the neutral pion with $\bar uu$ flavor
content for the smallest values of $N_b$. The plot shows that a chiral
extrapolation using the larger pion masses would lead to similar results.
However, in the region of small pion masses we observe deviations from the
straight line, possibly due to the presence of chiral logarithms. We show the
behavior of $C_u(a,B)$ in Fig.~\ref{fig:Z-vs-B}. Toward the continuum limit, 
$C_u$ approaches unity as is clearly visible in the
plot. In addition, for large values of $B$, we expect $C_u$ to approach the free fermion case
for which again $C_u\to1$. This is indicated by the increase in $C_u$ with increasing $B$. The
increase seems to level off at around $eB\approx2$~GeV$^2$, which could be a
sign for growing lattice artifacts.

The results for $\kappa_{c;u}$ and $\kappa_{c;d}$ are tabulated in
Table~\ref{tab:kappac}. To estimate the systematic uncertainty of the linear
ansatz within the present range of quark masses we have repeated the procedure by
including a term quadratic in $1/\kappa-1/\kappa_{c;f}$ in
Eq.~\refc{eq:kapc-extra}. The spread of the results from these two fits is
given as the systematic uncertainty in Table~\ref{tab:kappac}. To visualize the
effect of $B$ on $m_{c;f}$, we introduce the parameter
\be
 \label{eq:dm-def}
 \Delta m_{c;f}(a,B) = \frac{m_{c;f}(a,B) -
m_{c}(a,0)}{\mb_f(a,0;m_\pi=415\,{\rm MeV})} \,.
\ee
In this ratio and in our quenched setup, the leading $B=0$ lattice artifacts and
renormalization factors cancel, so this is the ideal tool to see to what extent
the renormalized quark mass is affected by the shift in $m_{c;f}(a,B)$. Note, that
if $\Delta m_{c;f}(a,B)=-1$, this means that the $B$-dependent part of the additive
quark mass renormalization is negative and as large in magnitude as the quark mass
associated with a pion with a mass of 415~MeV at $B=0$.

\begin{table}[b]
 \centering
 \begin{tabular}{c|cc|cc}
  \hline
  \hline
  & \multicolumn{2}{c|}{$\beta=5.845$} & \multicolumn{2}{c}{$\beta=6.000$} \\
  $N_b$ & $\kappa_{c;u}$ & $\kappa_{c;d}$ & $\kappa_{c;u}$ & $\kappa_{c;d}$ \\
  \hline
  1 & 0.16125(3)(4) & 0.16138(3)(12) & 0.15683(1)(3) & 0.156953(6)(34) \\
  2 & 0.16121(2)(4) & 0.16116(4)(8) & 0.15679(2)(2) & 0.156836(8)(8) \\
  3 & --- & --- & 0.157022(8)(15) & 0.156789(9)(28) \\
  4 & 0.16248(3)(1) & 0.16107(5)(6) & 0.157401(8)(17) & 0.156812(15)(18) \\
  6 & 0.16450(2)(2) & 0.16171(4)(2) & 0.158405(5)(9) & 0.157037(8)(16) \\
  8 & 0.16687(3)(2) & 0.16247(2)(2) & 0.159582(5)(24) & 0.157399(8)(26) \\
  10 & 0.16956(3)(1) & 0.16344(2)(2) & 0.160884(7)(30) & 0.157854(10)(15) \\
  12 & --- & --- & 0.162305(9)(25) & 0.158388(9)(12) \\
  \hline
  \hline
   &
  \multicolumn{2}{c|}{$\beta=6.260$} & & \\
  & $\kappa_{c;u}$ & $\kappa_{c;d}$ & & \\
  \hline
  1 & 0.152258(10)(3) & 0.152295(7)(17) & & \\
  2 & 0.152231(8)(23) & 0.152262(9)(14) & & \\
  4 & 0.152428(6)(22) & 0.152242(7)(8) & & \\
  6 & 0.152790(8)(13) & 0.152251(6)(23) & & \\
  8 & 0.153235(10)(21) & 0.152427(8)(23) & & \\
  10 & 0.153733(6)(11) & 0.152602(5)(8) & & \\
  12 & 0.154273(5)(13) & 0.152797(4)(3) & & \\
  \hline
  \hline
 \end{tabular}
 \caption{\label{tab:kappac}Results for $\kappa_{c;u}$ and $\kappa_{c;d}$
 obtained from the linear chiral extrapolation of the current quark masses.
 The first uncertainty is statistical, and the second is the estimate
 for the systematic uncertainty obtained from the spread to the
 result from an extrapolation with a polynomial of second order.}
\end{table}

\begin{figure}[t]
\begin{minipage}[c]{.47\textwidth}
 \includegraphics[]{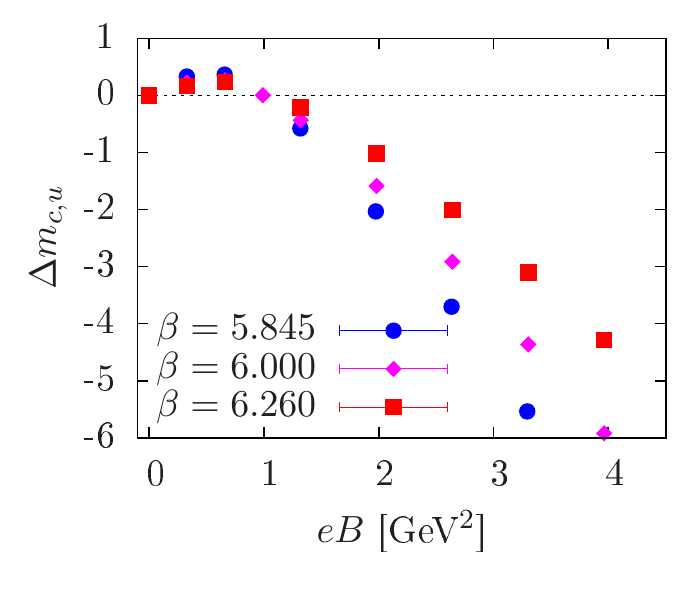}
\end{minipage}
\begin{minipage}[c]{.47\textwidth}
 \includegraphics[]{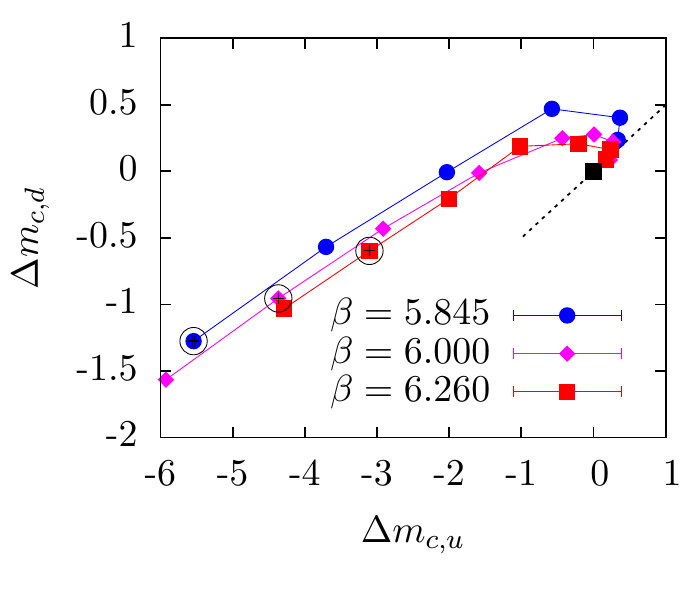}
\end{minipage}
\caption{Left: Results for $\Delta m_{c;u}(a,B)$ as defined in
Eq.~\refc{eq:dm-def} versus $B$ for different values of the lattice spacing.
Right: Results for $\Delta m_{c;f}(a,B)$ for different values of $B$ (starting
from the black point -- $B=0$ -- outward) and different lattice spacings. 
The black dashed line corresponds to the slope of 2,
which we expect for $B\approx0$ due to Eq.~\refc{eq:der-rel}, and the colored
lines simply connect the data points and are included to guide the eye. The
points marked with black circles correspond to a fixed physical magnetic field of
$eB=3.3$ GeV$^2$.
}
\label{fig:dm-plane}
\end{figure}

The results for $\Delta m_{c;u}$ and $\Delta m_{c;d}$ are displayed for
different values of $B$ and the different ensembles in Fig.~\ref{fig:dm-plane}.
In the left panel of the figure, we show $\Delta m_{c;u}$ versus $eB$ for the
different lattice spacings [$\Delta m_{c;d}$ follows from these results and 
Eq.~\refc{eq:der-rel}]. The plot indicates that the mass shift is positive for low 
magnetic fields, in contrast
to the negative shift in the free case [see Eq.~\refc{eq:free_tuning}].
However, the effects of higher-order terms in $B$ soon become important and
$\Delta m_{c;u}(a,B)$ becomes negative and linear in $B$ for strong magnetic
fields -- again in accordance 
with the expectation that we approach the free case in this limit.
The figure also shows that the magnitude of
$\Delta m_{c;u}(a,B)$ decreases with decreasing lattice spacing -- reflecting
the fact that the magnetic field dependence of $m_{c;f}$ is a lattice artifact and thus
vanishes in the continuum limit. The right panel displays the results in the
$(\Delta m_{u},\Delta m_{d})$-plane, and the individual data points indicate
different values of $B$, starting from $B=0$ (solid black square) and proceeding along
the colored lines which are included to guide the eye.

The snail shell-alike shape of the curves mapped out in
this parameter space demands some discussion. First, all curves start with a
slope which is close to 2 (indicated by the dashed black line). This is
expected due to the relation given in Eq.~\refc{eq:der-rel}, which fixes the
factor between the derivatives of $m_{c;u}$ and $m_{c;d}$ at $B=0$. The
direction of movement in the plane, however, is determined by the prefactor of
the first term appearing in the expansion of $m_{c;f}(a,B)$, Eq.~\refc{eq:mc-Bdep},
with respect to $B$
and we have seen above that it is positive. Deviations from the black line
indicate the onset of higher-order terms in the expansion of $m_{c;f}(a,B)$.
For large magnetic
fields we expect to approach the free theory (see Sec.~\ref{contiB}), where
the slope is again 2, cf.\ Appendix~\ref{app1}. This behavior is also visible
in Fig.~\ref{fig:dm-plane}. Next, let us discuss the change in shape for
$a\to0$. While the slopes for $B\to0$ and $B\to\infty$ are dictated by
Eq.~\refc{eq:der-rel} and the free theory case, respectively, the extent of the
curve depends on the lattice spacing. Remember that the $B$-dependence of
$m_{c;f}(a,B)$ is a lattice artifact, so the coefficients of the expansion
of $m_{c;f}(a,B)$ become smaller for $a\to0$, as can be seen in the left panel
of Fig.~\ref{fig:dm-plane}. Consequently, the movement along the curve as a
function of $B$ becomes slower, and in the continuum limit, the curve will
collapse to the starting point. To see this, compare the points of the curves
marked with the black circles -- these all correspond to the same physical
magnetic field $eB=3.302$~GeV$^2$ ($N_b=10$).

Using the results for $\kappa_{c;u}$ and $\kappa_{c;d}$ tabulated in
Table~\ref{tab:kappac}, we can now tune $\kappa_u$ and $\kappa_d$ along the LCP(B), 
so that the additively renormalized quark mass from Eq.~\refc{eq:bm-def} remains
independent of $B$.

\section{Quenched meson spectrum in external magnetic fields}
\label{sec:results}

\subsection{Computation of meson masses}
\label{theory-masses}

The meson masses are extracted from the large Euclidean-time behavior 
of the correlation functions according to
\be
\expv{O(t)O^\dagger(0)} = A \Big[ e^{-m_O(B) t} + e^{-m_O(B)(T-t)} \Big] ,
\label{eq:fitfunction}
\ee
where $O$ is either the pseudoscalar density (for the pion)
or the vector current (for the $\rho$-meson),
\be
\label{eq:vect-current}
P^a(x)=\bar\psi(x) \gamma_5 \tau^a \psi(x), \quad\quad 
V^a_\mu(x) = \bar{\psi}(x) \gamma_\mu \tau^a \psi(x) \,.
\ee
The charged mesons $a=\pm$ involve $\tau^\pm=(\tau^1\mp i\tau^2)/2$, while 
our (connected) neutral mesons either have pure $\bar u u$ [$\tau^u=\textmd{diag}(1,0)$] 
or pure $\bar d d$ [$\tau^d=\textmd{diag}(0,1)$] 
quark content.

In the presence of the background magnetic field, $\rho$-mesons
can be classified in terms of the projection of their spin onto the
magnetic field, $\ve{s}\ve{B}=\sz B$. In terms of the vector current
operator~\refc{eq:vect-current}, the spin eigenstates with $\sz=\pm1$ and 0 are given by
(see, e.g., Ref.~\cite{de1986field})
\be
\label{eq:spin-eigst}
V^a_\pm = \frac{1}{\sqrt{2}} \big( V^a_x \mp i V^a_y \big) \quad \text{and}
\quad V^a_{0} = V^a_z \,,
\ee
respectively. Thus, we can compute the relevant correlation functions for
$\sz=\pm1$ via
\be
\label{eq:rho-corrs}
C_{\rho^a,\pm}(t) = \frac{1}{2} \Big[ \expv{V^a_x(t)\big(V^a_x(0)\big)^\dagger}
+ \expv{V^a_y(t)\big(V^a_y(0)\big)^\dagger} \pm i \big(
\expv{V^a_x(t)\big(V^a_y(0)\big)^\dagger} -
\expv{V^a_y(t)\big(V^a_x(0)\big)^\dagger} \big) \Big]
\ee
and for $\sz=0$ by
\be
\label{eq:rho-0corr}
C_{\rho^a,0}(t) = \expv{V^a_z(t)\big(V^a_z(0)\big)^\dagger} \,.
\ee 
We remark that charge conjugation symmetry and parity symmetry 
ensure that $m_{\pi^+}=m_{\pi^-}$ and $m_{\rho^+_0}=m_{\rho^-_0}$ for the
modes with spin projection $s_z=0$ and 
$m_{\rho^+_+}=m_{\rho^-_-}$ and $m_{\rho^+_-}=m_{\rho^-_+}$ for the states
with nonvanishing spin projection on the $B$-field axis.

To determine the fit range for the function~\refc{eq:fitfunction}, we have
looked for the region of the minimal value of $t$ included in the fit, $t_{\rm
min}$, where $\chi^2/$dof is close to unity and where the extracted meson mass
does not depend on the particular choice of $t_{\rm min}$. For mesons with
masses larger than the neutral pion mass $m^{u}/m^{d}(B)$ the correlation
functions show the typical exponential decrease of the signal-to-noise ratio, so
that the maximal possible value for $t_{\rm min}$ decreases. In those cases we
have less control over contaminations from excited states and we note that these
effects should be further investigated in the future.

\subsection{Meson masses from lines of constant physics}
\label{sec:mlcp-masses}

\begin{figure}[t]
\begin{minipage}[c]{.47\textwidth}
 \includegraphics[]{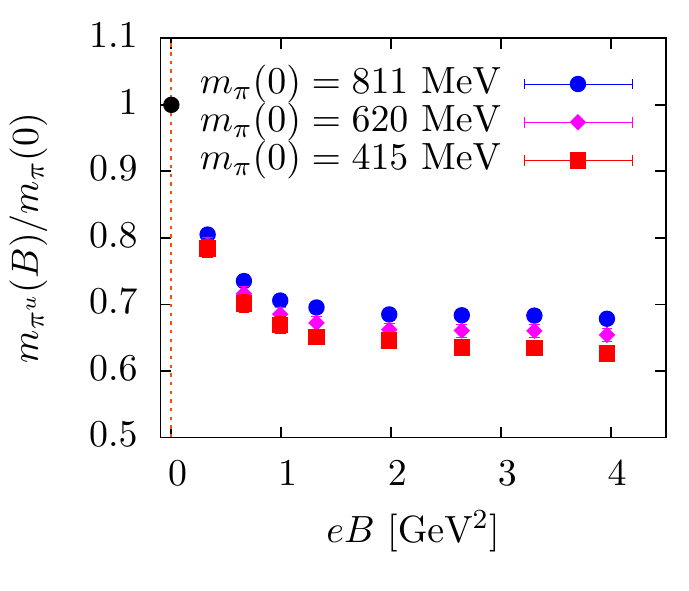}
\end{minipage}
\begin{minipage}[c]{.47\textwidth}
 \includegraphics[]{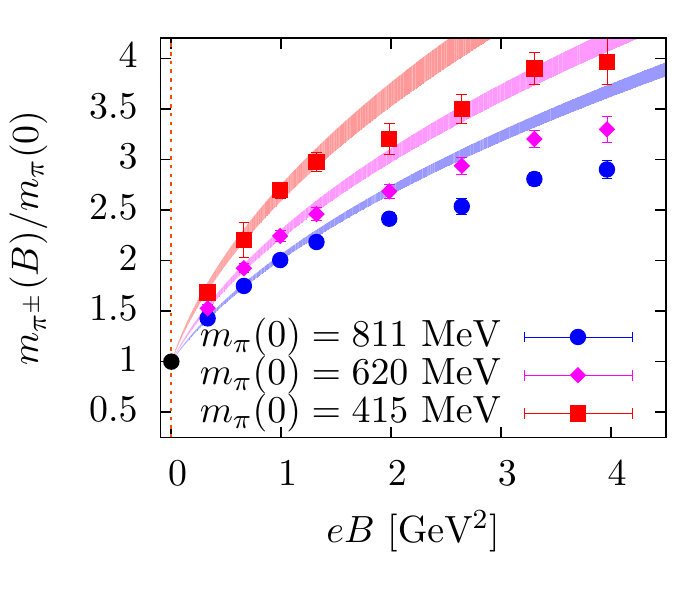}
\end{minipage}
\vspace*{-.4cm}
\caption{Results for the neutral ($\bar{u}u$) pion masses (left) and the masses
of $\pi^+$ (right) versus $B$ for different values of the quark masses on the
ensemble with $a=0.093$~fm. The filled curves correspond to the $\pi^+$ masses
in the free case, Eq.~\refc{eq:Escalar}.}
\label{fig:pions-qmdep}
\end{figure}

We have measured the meson masses along the LCP(B) using the tuned values of
$\kappa_{c;u}$ and $\kappa_{c;d}$ described in Sec.~\ref{sec:lcpB}. We 
work with quark masses corresponding to pion masses which are above
415~MeV at $B=0$, so that $m_\pi L\gtrsim3$ at $B=0$. In
this case, finite size effects for the meson spectrum are expected to be
small. A study of finite size effects at $B\neq0$ is presented in Sec.~\ref{fse}.
We start the discussion of the results
with the quark mass dependence of the meson masses on the 
$\beta=6.0$ ensemble, i.e., at the intermediate lattice spacing $a=0.093$~fm.

\begin{figure}[ht!]
\begin{minipage}[c]{.47\textwidth}
 \includegraphics[]{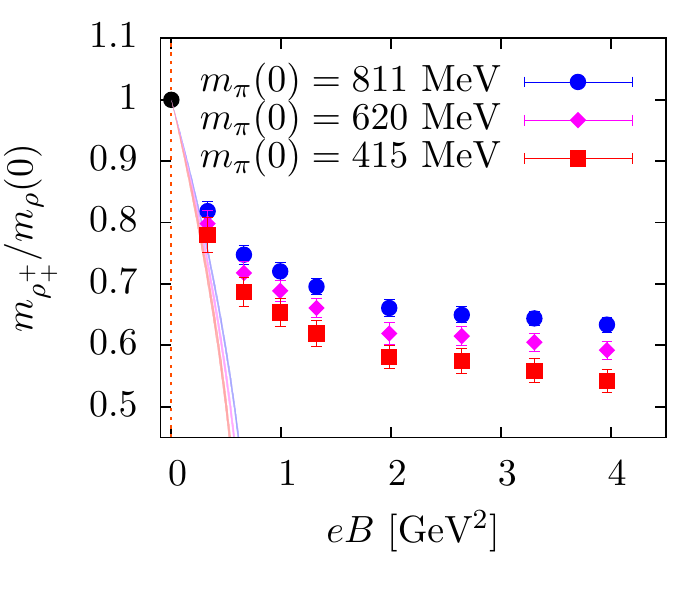}
\end{minipage}
\begin{minipage}[c]{.47\textwidth}
 \includegraphics[]{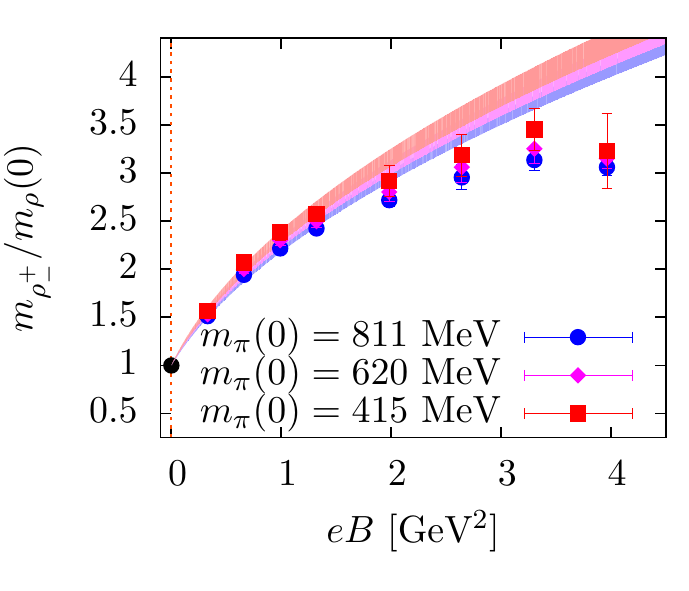}
\end{minipage}
\vspace*{-.2cm}
\caption{Results for $\rho^+$-meson masses with $\sz=+1$ (left) and $\sz=-1$
(right) versus $B$ for different values of the quark masses on the ensemble with
$a=0.093$~fm. The curves correspond to the masses in the free case,
Eq.~\refc{eq:enes-spin}.}
\label{fig:rhoc-mdep}
\end{figure}

In Fig.~\ref{fig:pions-qmdep}, we show the masses of neutral (left) and charged
pions (right) versus $B$, normalized by the mass at $B=0$.
The neutral pion masses fall off strongly before they
level off at intermediate values of $B$, decreasing monotonously throughout. 
At around $eB\approx 2$~GeV$^2$, the mass of the neutral pions has decreased
to about 60 to 70\% of its $B=0$ value. The quark mass dependence of this
behavior is rather mild but the effect is enhanced toward lighter
quark masses. The charged pions show the increase with the magnetic
field [cf.\ Eq.~\refc{eq:LLb} for noninteracting pions], 
with the tendency to undershoot the free-case prediction
for larger values of $B$. The same behavior has also been observed for
dynamical staggered quarks~\cite{Bali:2011qj} and 
perturbatively using one-loop pion-nucleon
interactions~\cite{Colucci:2013zoa}. 

\begin{figure}[t]
\centering
\begin{minipage}[c]{.47\textwidth}
 \includegraphics[]{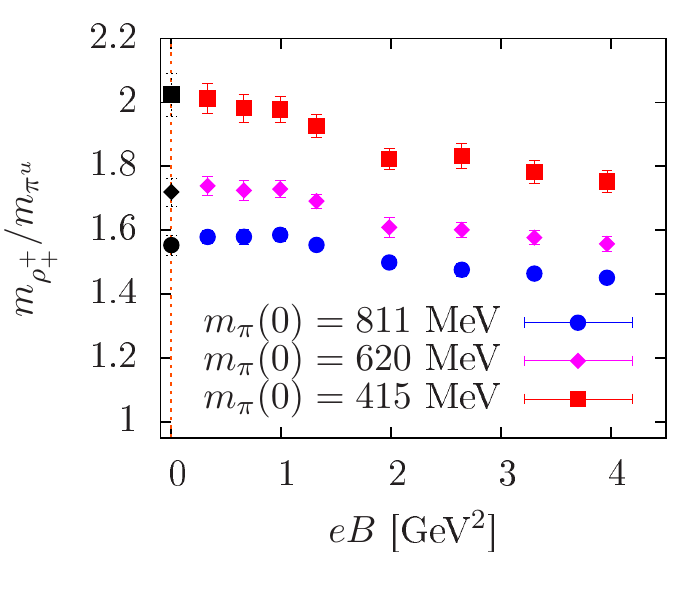}
\end{minipage}
\begin{minipage}[c]{.47\textwidth}
 \includegraphics[]{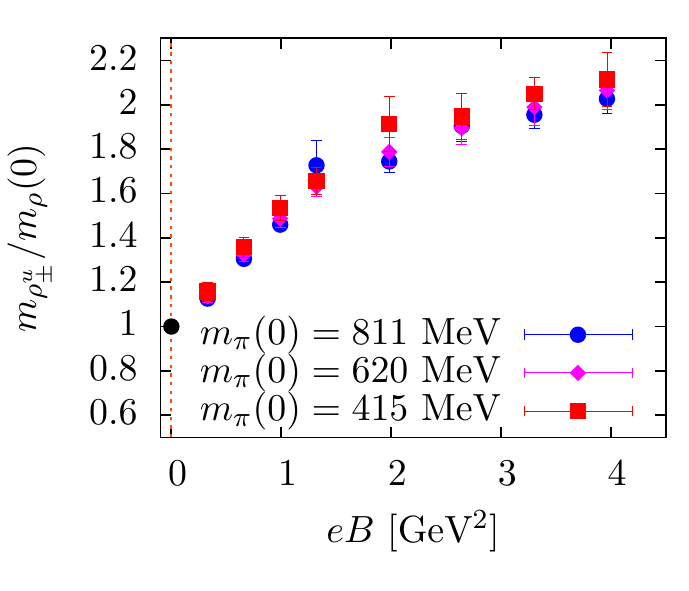}
\end{minipage}
\vspace*{-.2cm}
\caption{Left: Results for the $\rho^+$-meson masses with $\sz=+1$ versus $B$,
normalized to the neutral ($\bar{u}u$) pion mass at finite values of $B$.
Right: Results for $\rho^u$-meson masses with with $\sz=\pm1$ versus
$B$ for different values of the quark masses on the ensemble with $a=0.093$~fm.}
\label{fig:rhon-mdep}
\end{figure}

Next, we investigate $\rho$-mesons. Of particular relevance is the component $\rho^+_+$
(or equivalently the $\rho^-_-$), which is the particle conjectured to
condense~\cite{Chernodub:2011mc} for large values of $B$. First of all, we have checked that
our correlators are well described by the exponential form~\refc{eq:fitfunction} even
for the largest magnetic field $eB/m^2_{\rho}(0) \approx 6.7$.\footnote{Note that this is in
contrast with the quenched results of Ref.~\cite{Braguta:2011hq}; however, this reference uses much
smaller lattice extents. Note furthermore that Ref.~\cite{Braguta:2011hq} employed
correlators in the spatial $z$-direction. Nevertheless, at $T = 0$ and a
magnetic field parallel to the $z$ axis, this is expected to be equivalent to
correlation functions in the $t$-direction.} The results for the masses of the
$\rho^+_+$-mesons (left) and the $\rho^+_-$ (right) are shown in
Fig.~\ref{fig:rhoc-mdep} versus $B$, again normalized by the $B=0$ mass.
The plot indicates that the mass of the $\rho^+_{+}$
decreases monotonously with $B$, but it starts to level off at around
$eB\approx2$~GeV$^2$ and remains in the region of 50 to 70\% of its $B=0$ value.
In terms of the inequality~\refc{eq:ineq-bound}, this behavior is
expected, given the flattening of $m_{\pi^u}$ and $m_{\pi^d}$ at
large $B$. In fact, the $\rho^+_{+}$-meson mass does not saturate the bound
from Eq.~\refc{eq:ineq-bound} but remains at a multiple of the pion mass larger
than 1, as shown in Fig.~\ref{fig:rhon-mdep} (left).
The plot indicates that the ratio $m_{\rho^+_+}(B)/m_{\pi^0}(B)$ remains
constant within 20\% with $B$ and increases as the pion becomes lighter. This
indicates that one moves further away from the saturation of the bound. The $\rho^+_{-}$
masses show the
increase expected from the free case, Eq.~\refc{eq:enes-spin} with $g=2$, but
tend to undershoot the free-case curve for larger magnetic fields.
Once more, this feature has only a very mild quark mass dependence.

In Fig.~\ref{fig:rhon-mdep} (right), we show the masses of
the neutral $\rho^u_{\pm}$-meson with nonzero $\sz$.
Since the neutral meson has no
magnetic moment, this mass remains unaffected by the magnetic field in the free
case, and indeed we find that the mass is insensitive to the spin direction.
However, we observe 
an almost linear rise with $B$, in
this case almost completely independent of the quark mass.
We suspect that 
disconnected diagrams might be important for this channel.

\begin{figure}[t]
\begin{minipage}[c]{.47\textwidth}
 \includegraphics[]{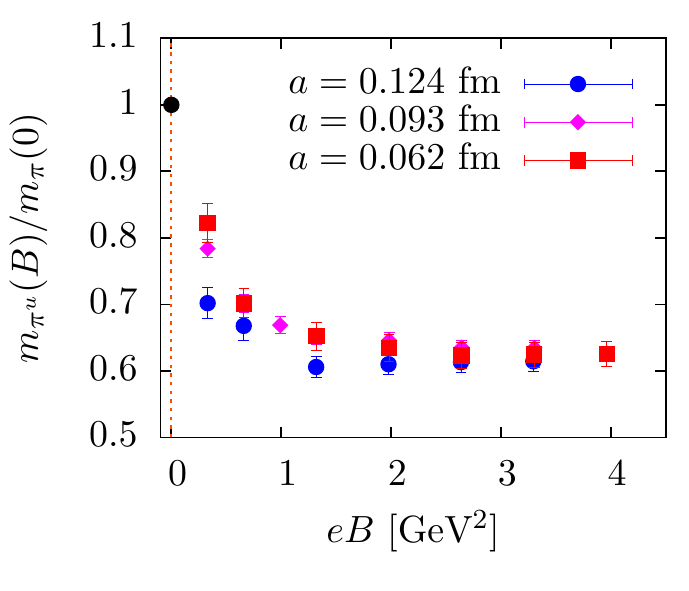}
\end{minipage}
\begin{minipage}[c]{.47\textwidth}
 \includegraphics[]{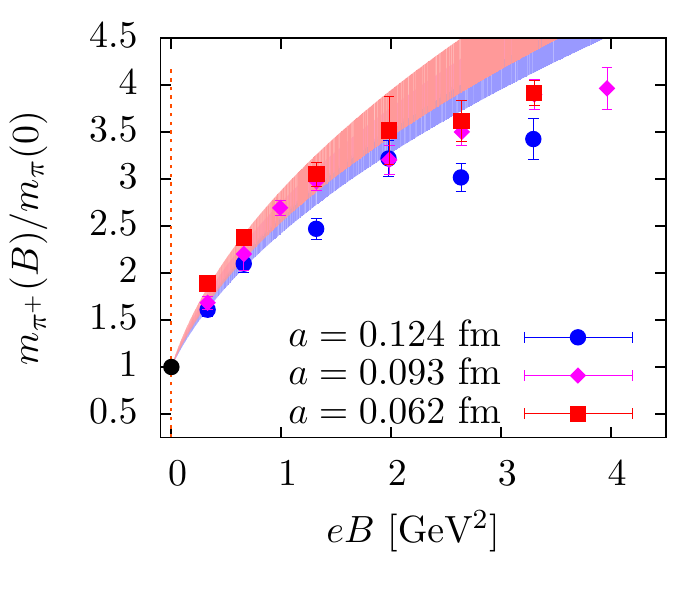}
\end{minipage}
\vspace*{-.2cm}
\caption{Results for the neutral ($\bar{\rm u}u$) pion masses (left) and the masses
of $\pi^+$ (right) versus $B$ for different values of the lattice spacing
for $m_\pi(0)=416$~MeV. The filled curves
correspond to the $\pi^+$ masses in the free case, Eq.~\refc{eq:Escalar}.}
\label{fig:pions-adep}
\end{figure}

\begin{figure}[ht!]
\begin{minipage}[c]{.47\textwidth}
 \includegraphics[]{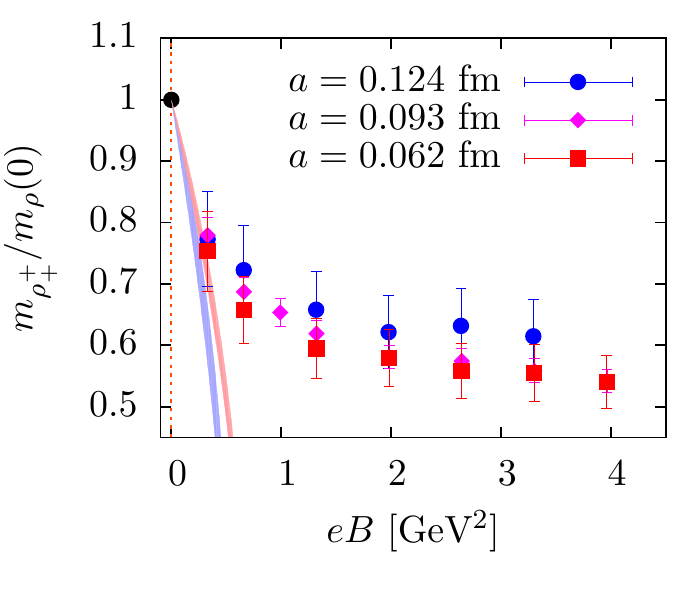}
\end{minipage}
\begin{minipage}[c]{.47\textwidth}
 \includegraphics[]{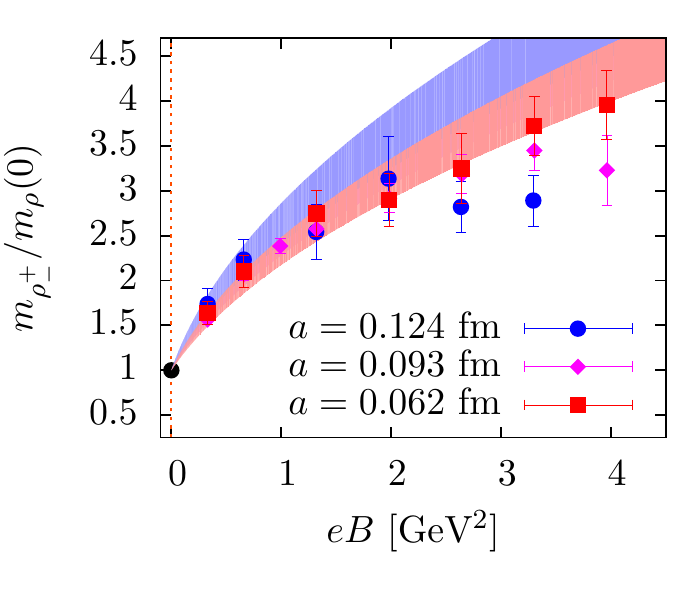}
\end{minipage}
\caption{Results for $\rho^+$-meson masses with $\sz=+1$ (left) and $\sz=-1$
(right) versus $B$ for different values of the lattice spacing for
$m_\pi(0)=416$~MeV. The curves correspond to
the masses in the free case, Eq.~\refc{eq:enes-spin}.}
\label{fig:rhoc-adep}
\end{figure}

\begin{figure}[ht!]
\begin{minipage}[c]{.47\textwidth}
 \includegraphics[]{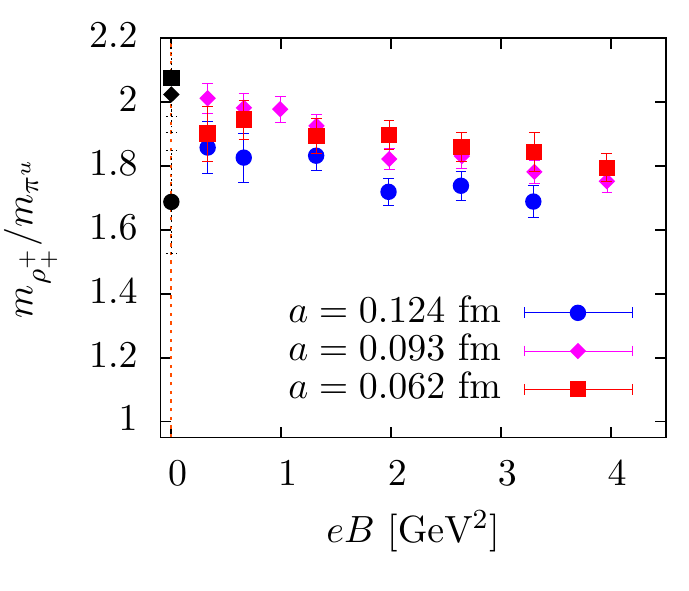}
\end{minipage}
\begin{minipage}[c]{.47\textwidth}
 \includegraphics[]{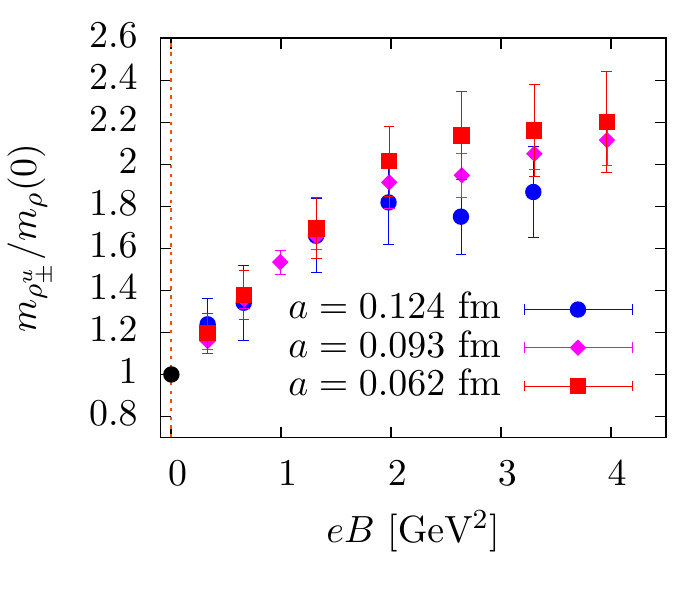}
\end{minipage}
\caption{Left: Results for the $\rho^+$-meson masses with $\sz=+1$ versus $B$,
normalized to the neutral pion mass at finite values of $B$ for $m_\pi(0)=416$~MeV.
Right: Results for $\rho^u$-meson masses with $\sz=\pm1$ versus
$B$ for different values of the lattice spacing for an approximately constant
pion mass of 416~MeV at $B=0$.}
\label{fig:rhon-adep}
\end{figure}

Next, we investigate the dependence of the meson masses on the lattice spacing
(the final continuum
extrapolation is postponed to Sec.~\ref{sec:conti-extr}). To this end we keep
the quark mass constant by fixing the pion mass at $B=0$ to 415~MeV. The results
for the neutral (left) and charged pions (right) versus $B$ for different values
of the lattice spacing are shown in Fig.~\ref{fig:pions-adep}. The plot
indicates that in both cases only mild lattice artifacts are present. 
The result of the same analysis for the $\rho^+_+$ and $\rho^+_-$-states
is shown in Fig.~\ref{fig:rhoc-adep} in
the left and right panels, respectively. As for the pion masses, 
we observe only a mild lattice spacing dependence.
Concerning the saturation of the bound~\refc{eq:ineq-bound},
reducing the lattice spacing
tends to drive the results away from saturation, as can be seen from
Fig.~\ref{fig:rhon-adep} (left). This means that cutoff effects and quark mass
effects act in the same manner in this respect.
The mass of the neutral $\rho^u_{\pm}$-meson
with nonzero $\sz$, shown in Fig.~\ref{fig:rhon-adep} (right), is also mainly
independent of the lattice spacing with the tendency to show noticeable effects
for larger values of $B$.

\subsection[Pion-$\rho$-meson mixing]{\boldmath Pion-$\rho$-meson mixing}

We will now discuss the mixing between pions and $\rho$-mesons with
spin projection $\sz=0$ along the magnetic field axis, as described in Sec.~\ref{sec:mixing}. 
Since we neglect
disconnected diagrams, we will focus on the mixing between the charged meson pair
$\pi^+$ and $\rho^+_0$.
Note that taking into account this mixing is necessary for the determination of 
the mass of the heavier of the pair (the $\rho^+_0$), but it does not affect the lighter 
state (the $\pi^+$).

Due to the mixing described in Eq.~\refc{eq:pi-rho-mix}, for $B>0$, 
the naive operators for the states
$\pi^+$ and $\rho^+_0$ overlap.
The orthogonal basis can be determined via the diagonalization of the
correlator matrix
\be
\begin{pmatrix}
 \langle \pi^+ \pi^+ \rangle & \langle \pi^+ \rho^+_0 \rangle \\
 \langle \rho^+_0 \pi^+ \rangle & \langle \rho^+_0 \rho^+_0 \rangle \\
\end{pmatrix} (t) \,,
\ee
and the true masses are the ones associated with the temporal exponential
decay of the eigenvalues of this matrix in the $t\to\infty$ limit. 
We remark that the off-diagonal 
elements need to be multiplied by $\pm i$ when we translate from the 
Euclidean to the Minkowskian Dirac matrices (for the spatial components
these differ by a factor $i$ so that $\gamma_z^{\rm E}=-i \gamma_z^{\rm M}$).
To extract the eigenvalues of the matrix, we use a generalized eigenvalue
problem (see Ref.~\cite{Blossier:2009kd} and references therein) with starting
temporal extents of $t_0=3a,\,4a$ and $5a$ for $\beta=5.845,\,6.0$ and $6.26$,
respectively. In the construction of the correlator matrix we have used
smeared operators at the source and sink, so the correlator matrix ought to be
symmetric. In practice, this only holds within the statistical uncertainty, and
we found it beneficial to stabilize the computation of eigenvalues by replacing
the off-diagonal elements by (the real part of) their average.

\begin{figure}[t]
\begin{minipage}[c]{.47\textwidth}
 \includegraphics[]{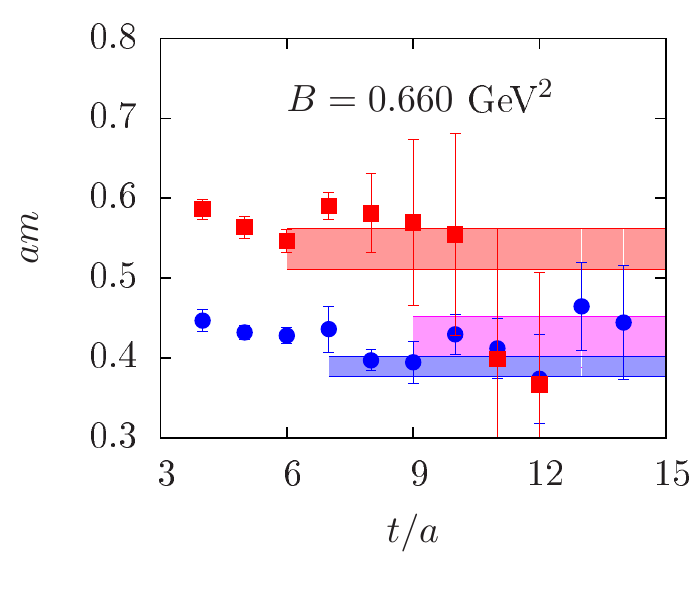}
\end{minipage}
\begin{minipage}[c]{.47\textwidth}
 \includegraphics[]{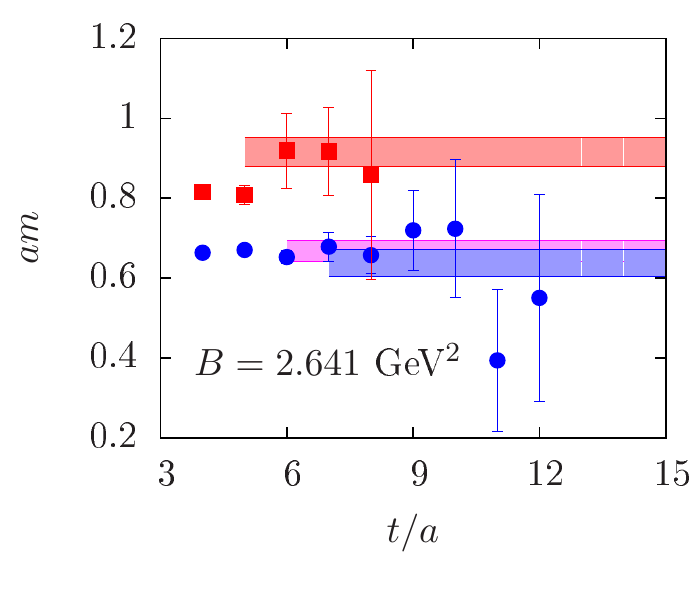}
\end{minipage}
\caption{Results for the effective masses for the two different eigenvalues
(blue circles for the ground state and red boxes for the first excited state)
obtained from the GEVP in the channel relevant for $\pi^+$ and $\rho^+_0$
mixing. The results have been obtained on the ensemble with $a=0.093$~fm
with a $B=0$ pion mass of 416~MeV. The colored areas are the results from
a fit to the correlation function. The magenta areas are the ones for
the analysis of the $\pi^+$ correlation function alone, without the use of
the GEVP.}
\label{fig:rho-mixing-effm}
\end{figure}

\begin{figure}[t]
\begin{minipage}[c]{.47\textwidth}
 \includegraphics[]{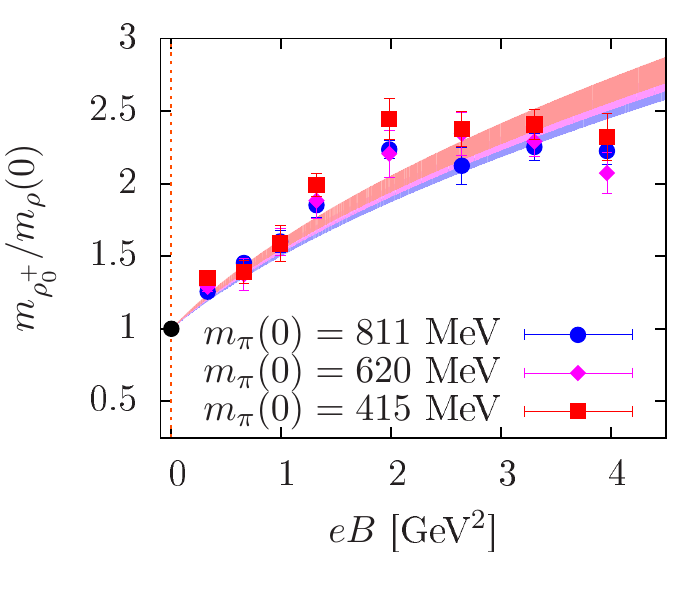}
\end{minipage}
\begin{minipage}[c]{.47\textwidth}
 \includegraphics[]{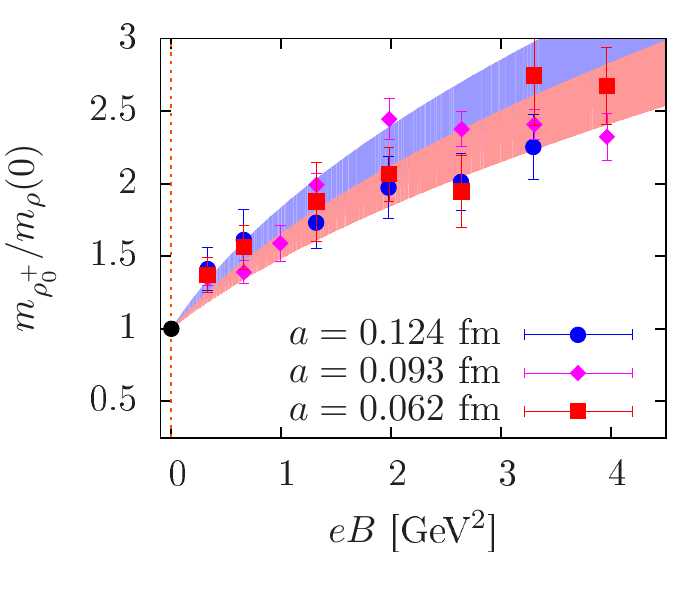}
\end{minipage}
\caption{Results for $\rho^{\pm}_{\sz=0}$ versus $B$ for different values of the
quark masses on the ensemble with $a=0.093$~fm (left) and for different values
of the lattice spacing at a constant $B=0$ pion mass of 416~MeV (right). The
curves correspond to the masses in the free case, Eq.~\refc{eq:enes-spin}.}
\label{fig:rho-mixing}
\end{figure}

The smaller eigenvalue of the correlator matrix corresponds to the 
heavier state, i.e., to the $\rho^+_{0}$-meson.
To extract the mass of the $\rho^+_{0}$-meson, we perform a fit
similar to that in Eq.~\refc{eq:fitfunction}. We show the associated effective
masses for the two eigenvalues for the $\beta=6.0$ ensemble for $eB=0.660$
and 2.641~GeV$^2$ in Fig.~\ref{fig:rho-mixing-effm}. Note, that the GEVP has been
setup with $t_0=4a$ and that the signal soon becomes lost in noise, so we
have left out the results for the effective masses once uncertainties become
overly large. For comparison, we have also plotted the result for $\pi^+$ from
the analysis of the pion correlator without the use of the GEVP. The results
indicate good agreement for the mass of the ground state in this channel.
The final masses for different values of the quark mass are shown in
Fig.~\ref{fig:rho-mixing} (left),
normalized to the $\rho$-meson mass at $B=0$.
We see that the mass of the $\rho^+_{0}$ increases and is in
agreement with the free case prediction, the colored lines, even though it shows
the tendency to overshoot the prediction for intermediate values of $B$ and to
undershoot for $B\gtrsim 3\textmd{ GeV}^2$. This agreement lends support to our
method and suggests that the contamination from other excited states is suppressed.
The dependence of the mass on the lattice spacing is shown in Fig.~\ref{fig:rho-mixing}
(right), revealing that, as before, lattice artifacts are smaller than our statistical
errors. Note once more, that we are working in the quenched approximation, where
sea quarks are absent and the $\rho$-meson is a stable particle. Unlike in the case with
dynamical fermions (e.g., Refs.~\cite{McNeile:2002fh,Bali:2015gji}), we thus do not
need to consider multipion states in our analysis.

\subsection{Results in the continuum limit}
\label{sec:conti-extr}

\begin{figure}[t]
\begin{minipage}[c]{.47\textwidth}
 \includegraphics[]{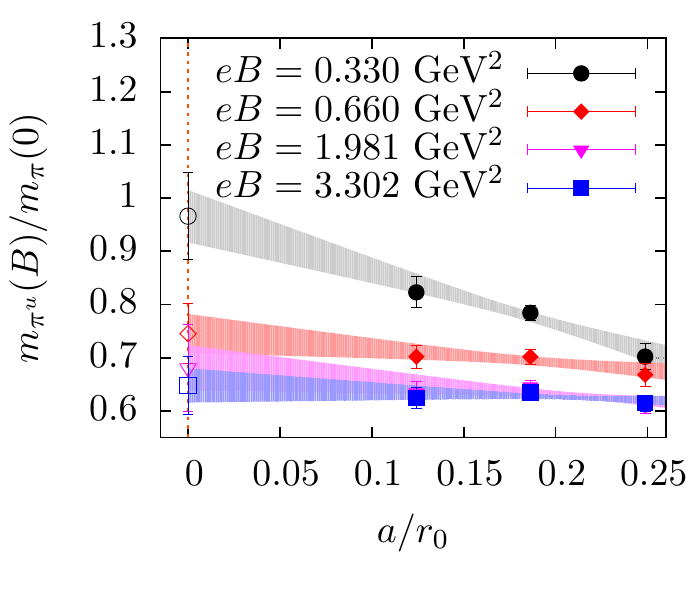}
\end{minipage}
\begin{minipage}[c]{.47\textwidth}
 \includegraphics[]{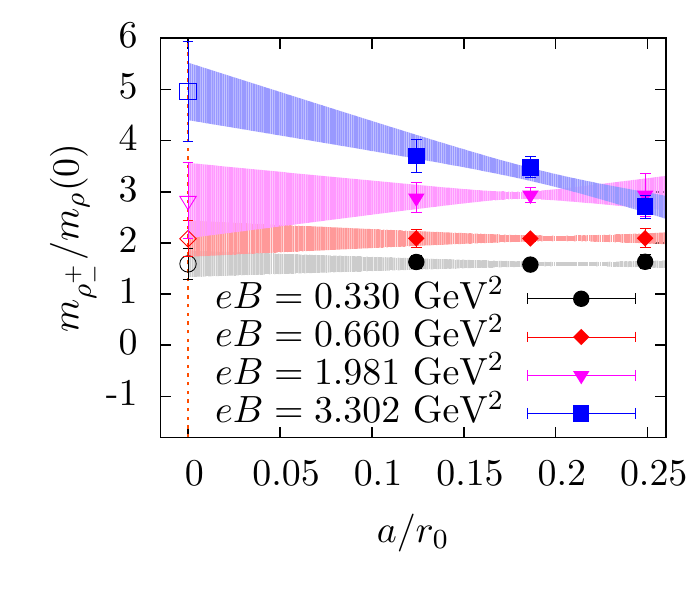}
\end{minipage}
\caption{Continuum extrapolations of the neutral ($uu$) pion
masses (left) and the $\rho^+_-$ masses for different values of $B$,
normalized by the mass at $B=0$. The filled curves represent linear
fits including all data points. The error bar at $a=0$ includes in addition 
the systematic error obtained by varying the fit range.}
\label{fig:conti-extrapol}
\end{figure}

We proceed by performing the continuum extrapolation of
the meson masses. As in the previous section we will work at a fixed $B=0$ pion
mass of 415 MeV. Since we employ unimproved Wilson fermions, meson
masses should show lattice artifacts of $\Ord(a)$. Consequently, our continuum
extrapolation is carried out linearly in $a$ and we extrapolate the meson masses for
each value of $B$ individually. The results for this continuum extrapolation are
shown for two representative cases [the neutral ($uu$) pion masses and the
$\rho^+_{-}$ masses] in Fig.~\ref{fig:conti-extrapol}. The linear
extrapolation works well for most of the cases, giving $\chi^2/$dof values in the
region between 0.5 and 2. However, we cannot exclude that our data still
receive significant contributions by terms of $\Ord(a^2)$. To estimate the
associated systematic uncertainty we have performed another linear continuum
extrapolation including only the points at the two smallest lattice spacings and
use the difference between the two extrapolations as the systematic uncertainty.
In Fig.~\ref{fig:conti-extrapol} the resulting uncertainty, including the
systematic part, is represented by the error bars of the points on the $a=0$
intersect. The continuum extrapolations at high magnetic field suggest 
that $O(a^2)$-effects become more pronounced as $B$ grows. In this region 
the systematic errors are potentially underestimated and should be checked
in future simulations on finer lattices.

\begin{figure}[t]
\begin{center}
 \includegraphics[]{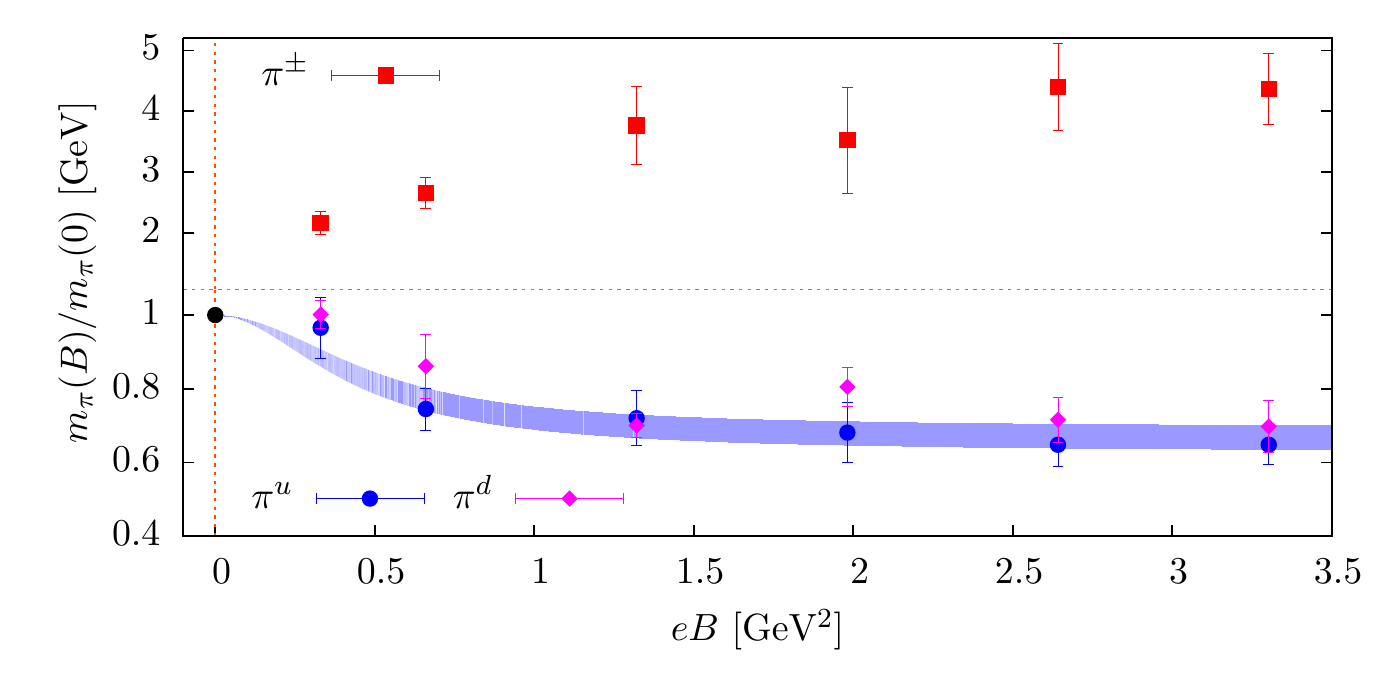} \\
 \includegraphics[]{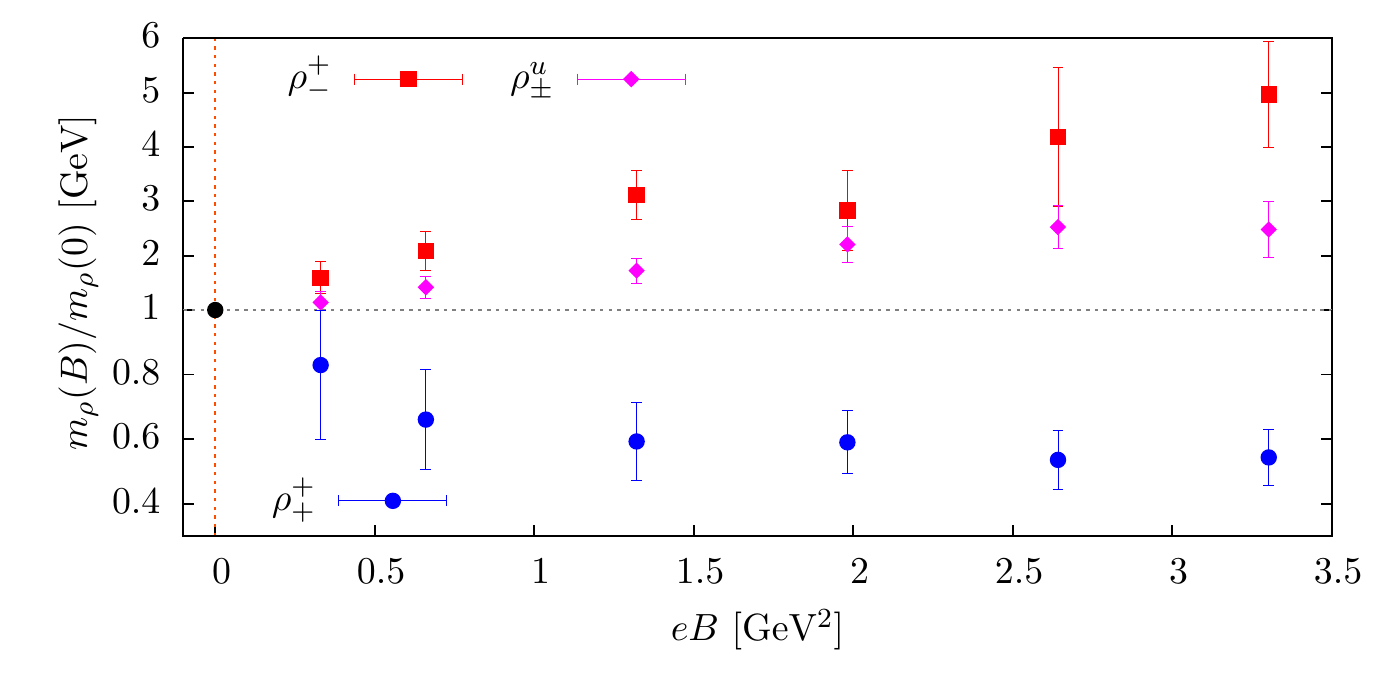}
\end{center}
\caption{Continuum results for pion (top) and $\rho$-meson (bottom) masses
versus $B$ for a $B=0$ pion mass of 415 MeV, normalized to the respective $B=0$
values. Note, that the scaling of the $y$ axes changes at the dashed horizontal
line.}
\label{fig:conti-mesons}
\end{figure}

We show the results for the continuum extrapolated meson masses in
Fig.~\ref{fig:conti-mesons}. The results basically confirm what we already found
at finite lattice spacing in the previous section. The masses of the neutral
pions (with $\bar{u}u$ and $\bar{d}d$ flavor content) decrease down to about
60\%-70\% of their $B=0$ value at $B\gtrsim2$~GeV$^2$, while the masses of the
charged pions increase, in agreement with the energies in the free case,
Eq.~\refc{eq:Escalar}. The data for $m_{\pi^u}$ can be qualitatively described by a
curve of the form
\be
\label{eq:emp-curve}
\frac{m_{\pi^u}(B)}{m_\pi(0)} = \frac{1+a_1(eB)^2}{1+a_2(eB)^2} \,.
\ee
Performing a fit to the data, including also the data for $m_{\pi^d}$, with $B\to B/2$
following Eq.~\refc{eq:der-rel}, we obtain $a_1=3.2(8)$ GeV$^{-2}$ and
$a_2=4.8(1.2)$ GeV$^{-2}$. The resulting curve is shown in
Fig.~\ref{fig:conti-mesons} (top), too.

The mass of the $\rho^+_+$, similarly to that of the neutral pion,
decreases down to about 60\% of its $B=0$ value, where it starts to level off.
In fact, the $\rho^+_+$ mass remains at twice the mass of the $\pi^u$.
The masses of the other $\rho$-mesons
with $\sz=\pm1$ increase, even those of the $\rho^{u/d}_{\pm}$-mesons, the
$B$-dependence of which is, as for the neutral pions, only an indirect effect of $B$,
due to the nonvanishing polarizability of the meson.

\subsection[Comparison to trajectories with $B$-independent $\kappa$]{Comparison to trajectories with \boldmath$B$-independent $\kappa$}
\label{compare}

In the previous sections, we have discussed the results for meson masses obtained
along the LCP(B), where the bare quark mass has been tuned in order for the 
renormalized quark mass to remain constant with
varying $B$. In the literature, this tuning has not been considered so far, and we will
now show the problems that can arise without it. Of course, results obtained on
the LCP(0) should also give the correct continuum limit; however, the lattice
artifacts in this case strongly depend on $B$, as we will demonstrate.
In this section we will focus on the results for the neutral
pion with $uu$ flavor content, since these are the ones which are affected
most by changes in the quark mass. However, other meson masses are influenced similarly 
by this effect.

\begin{figure}[t]
\begin{center}
 \includegraphics[]{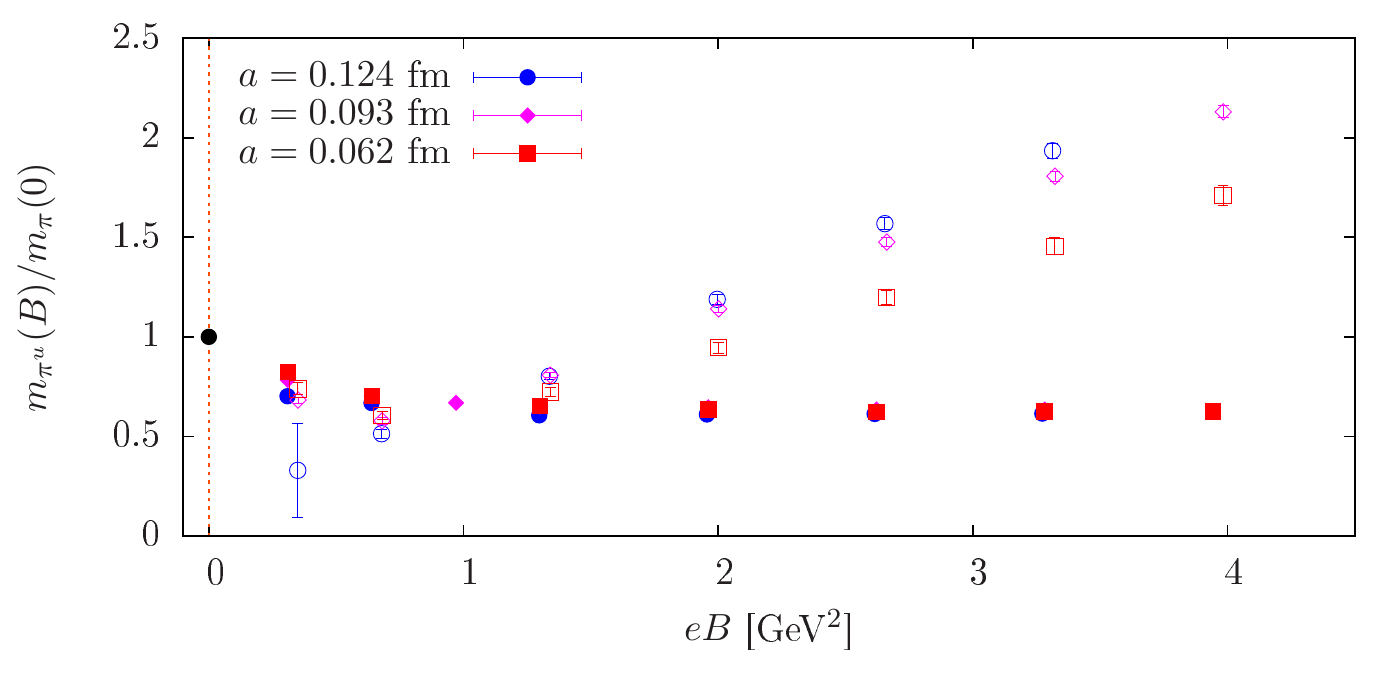}
\end{center}
\caption{Results for the neutral ($\bar{u}u$) pion masses versus $B$ for a pion mass
of 415~MeV at $B=0$ and different values of the lattice spacing, normalized to
their $B=0$ value. The open symbols correspond to results obtained with constant
$\kappa$, while the filled symbols correspond to the LCP(B). We have
slightly shifted the results horizontally for better visibility of the different
data sets.}
\label{fig:compare-pions}
\end{figure}

\begin{figure}[t]
\begin{minipage}[c]{.47\textwidth}
 \includegraphics[]{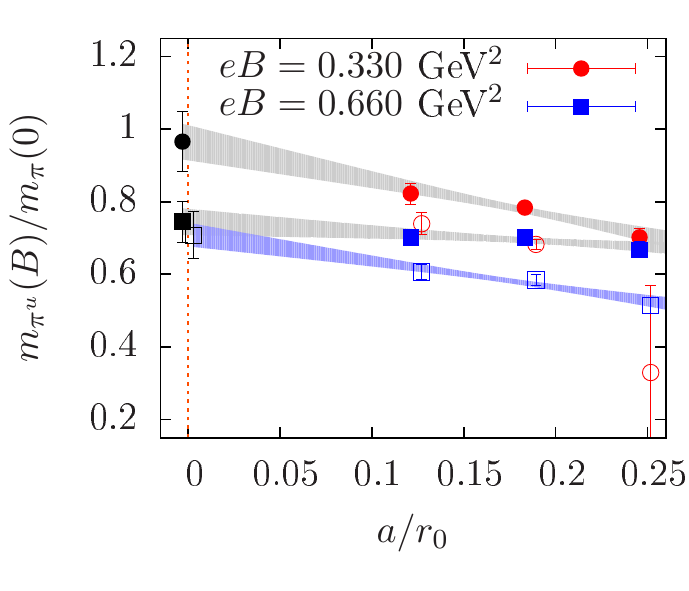}
\end{minipage}
\begin{minipage}[c]{.47\textwidth}
 \includegraphics[]{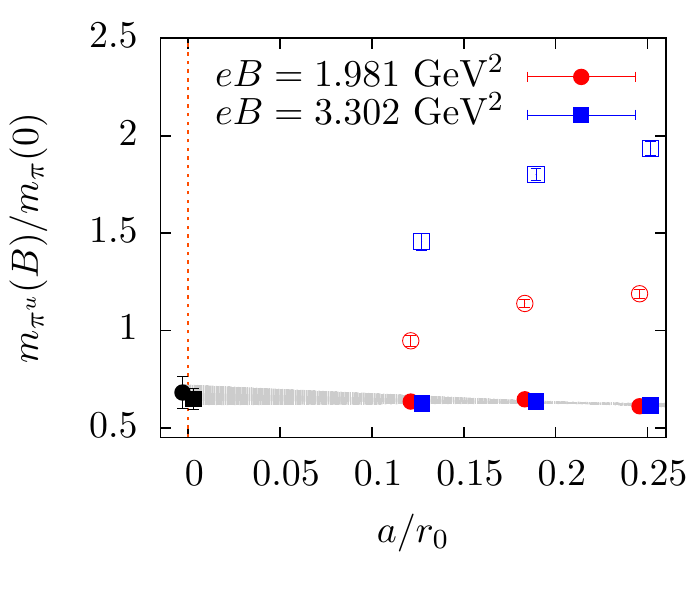}
\end{minipage}
\caption{Results for the masses of the neutral ($\bar{u}u$) pions versus the lattice
spacing in units of $r_0$, normalized by the mass at $B=0$. The filled points
correspond to the results obtained along the LCP(B) and the filled gray curves represent
their continuum extrapolation. The open symbols denote the
results obtained without $B$-dependent improvement. In this case, a continuum extrapolation 
is only possible for $eB=0.66 \textmd{ GeV}^2$ (blue curve). 
The black points on the $a=0$ line, indicated by the vertical dashed line, are the results
from the continuum extrapolations, including the estimate for systematic uncertainties.
In the figures we have slightly shifted some of the results horizontally (including the
associated curves) for better visibility of the different data sets.}
\label{fig:conti-extrapol2}
\end{figure}

In Fig.~\ref{fig:compare-pions}, we show the results for the masses
obtained with constant $\kappa$ for different values of $B$ compared to those
calculated along the LCP(B) for different lattice spacings,
corresponding to a $B=0$ pion mass of 415~MeV. The plot indicates that the
results with constant $\kappa$ suffer from enormous lattice artifacts. 
These act in a way in which the meson masses are initially below the continuum result, 
while they overshoot it for strong $B$. This tendency originates from the
nonmonotonous behavior shown in Fig.~\ref{fig:dm-plane}, where the critical mass
initially becomes bigger, before it starts to decrease with $B$.
To investigate the different lattice artifacts in more detail,
we plot the masses against the lattice spacing for some representative cases in
Fig.~\ref{fig:conti-extrapol2}. The plots show that the two sets of results seem to converge 
toward each other as $a\to0$. However, in most cases, a controlled (linear) 
continuum extrapolation is not possible for the results at constant $\kappa$. 
The exception is for $eB=0.660$~GeV$^2$, where both linear extrapolations point to 
the same continuum limit. Summarizing, the
results for constant $\kappa$ show lattice artifacts that are much larger than
those of the runs along the LCP(B), which is particularly true for large values of $B$.

\begin{figure}[t]
\begin{center}
 \includegraphics[]{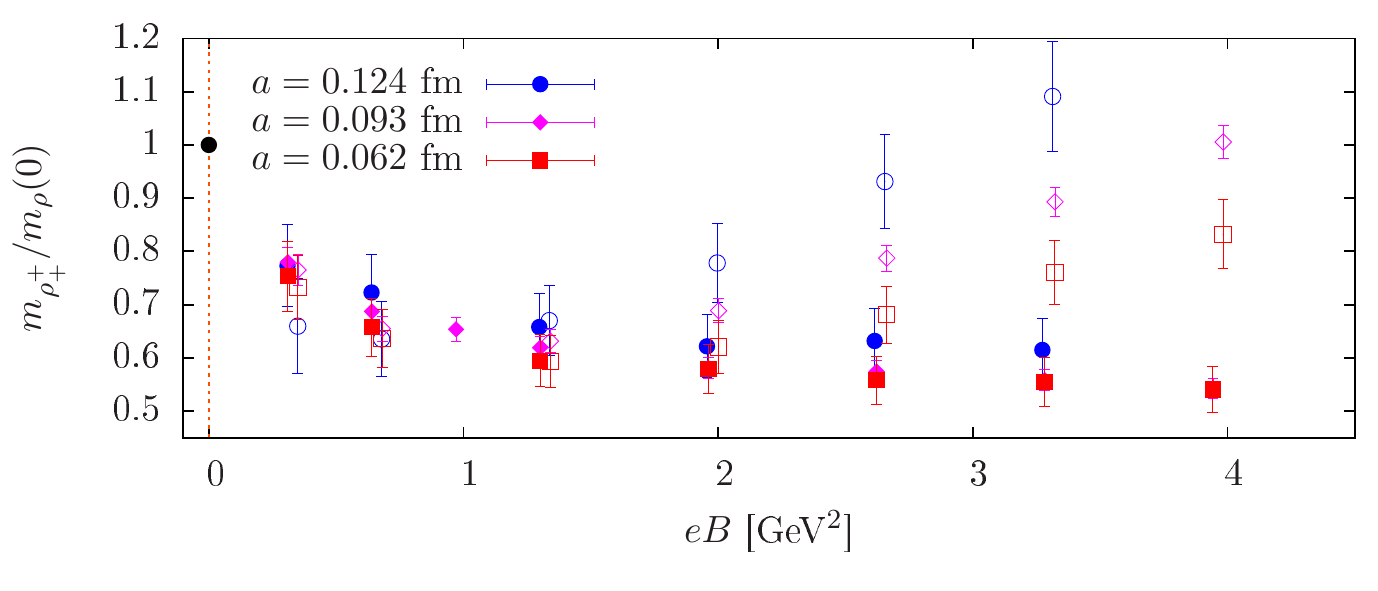}
\end{center}
\caption{Results for the $\rho^+_{+}$-meson masses versus $B$ for a pion
mass of 415~MeV at $B=0$ and different values of the lattice spacing, normalized
to their $B=0$ value. The open symbols correspond to results obtained with
constant $\kappa$, while the filled symbols show results along the LCP(B). We have
slightly shifted the results horizontally for better visibility of the different
data sets.}
\label{fig:compare-rhos}
\end{figure}

In Fig.~\ref{fig:compare-rhos}, we compare the
$\rho^+_+$-meson mass results obtained from constant $\kappa$ and from
the LCP(B), revealing the same tendency as observed above for the pions.

\subsection{Finite size effects}
\label{fse}

In the previous sections, we have seen that the neutral pion mass decreases down
to 60\% of its $B=0$ value with increasing $B$. Consequently, finite
size effects will potentially increase, since $m_{\pi^u}(B) L$ becomes smaller.
Indeed, for the strongest magnetic field, 
this combination drops to a value of $m_{\pi^u}(B) L\approx2$, indicating that finite size effects
can potentially be sizable. To investigate this systematic uncertainty, we
have generated configurations on a $48\times24^3$ lattice at $\beta=6.0$,
providing a second, larger volume for this $\beta$ value compared to the
$48\times16^3$ lattice listed in Table~\ref{tab:runp}, which has been used
in the previous sections. On these configurations, we have computed the meson
masses for $eB=1.320$, $2.641$, and $3.962\textmd{ GeV}^2$ -- these magnetic fields 
can be realized with integer flux quanta $N_b$ on both volumes; see
Eq.~\refc{eq:fquant}. For the computations, we have used the $\kappa$ values
obtained from the LCP(B) determined on the smaller volume. 

\begin{figure}[ht!]
\begin{minipage}[c]{.47\textwidth}
 \includegraphics[]{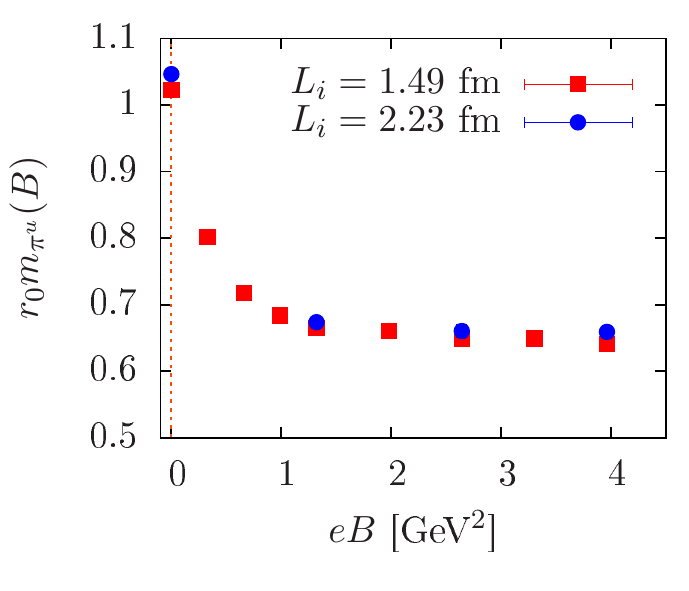}
\end{minipage}
\begin{minipage}[c]{.47\textwidth}
 \includegraphics[]{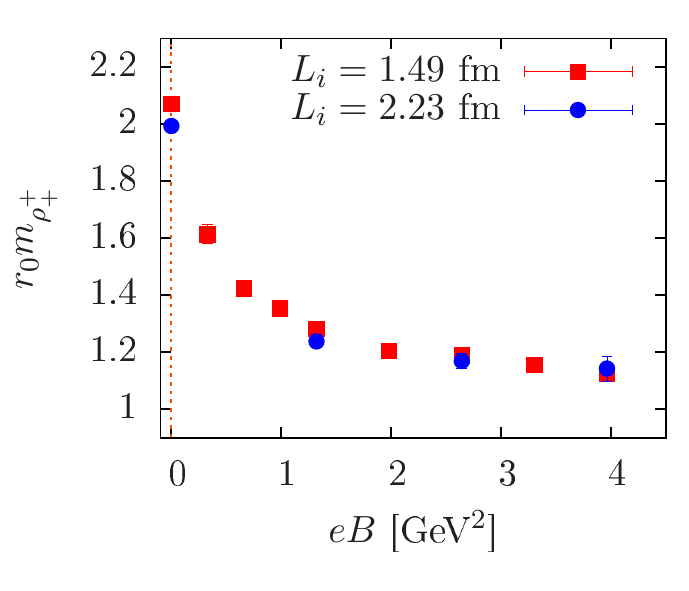}
\end{minipage}
\caption{Results for the masses of the neutral ($\bar{u}u$) pions (left)
and the $\rho_+^+$-mesons (right) in units of the vacuum Sommer parameter 
$r_0=0.5 \textmd{ fm}$. Shown are two different spatial volumes of
$16^3$ and $24^3$ at $\beta=6.0$, corresponding to $m_{\pi^u}(B) L\approx2$ and
3, respectively, for the largest value of $B$.}
\label{fig:fse-mesons}
\end{figure}

The results for the two different volumes are shown in Fig.~\ref{fig:fse-mesons}
for the neutral pion and the $\rho_+^+$, which are the two most important mesons
for our analysis.
The plot indicates that finite size effects are small.
For both mesons, the results change by around 2\%-3\% which is well below
the typical uncertainties of more than 10\% of our continuum extrapolated results.
The same is true for the other meson masses. 
Thus, we conclude that increasing $B$ does not enhance
finite size effects and that these at present can be neglected in
comparison to other systematics.

\section{Background field method and polarizabilities with Wilson fermions}
\label{polar}

\subsection{Background field method and additive quark mass renormalization}

Further interesting features of hadrons are their magnetic moments and the electric and
magnetic polarizabilities. The latter describe the indirect response of a
bound state to external electromagnetic fields. The background field method gives
access to both the magnetic~\cite{Martinelli:1982cb} and the electric
(see Refs.~\cite{Fiebig:1988en,Christensen:2004ca} for instance) properties.
In fact, in the past few years, a
number of groups have started to measure these quantities in lattice QCD
using Wilson fermions;
see Refs.~\cite{Lee:2005dq,Detmold:2006vu,Lee:2008qf,Detmold:2009dx,Detmold:2010ts,Primer:2013pva,Hall:2013dva,Lujan:2014kia,Freeman:2014kka,Beane:2014ora,Chang:2015qxa,Lujan:2016ffj}.
So far, the change of the additive renormalization of the quark mass has been neglected
in these studies, each of which were carried out at a single lattice spacing.
Before investigating the magnitude of this effect, let us first discuss the
implications that the unwanted change in the renormalized quark mass in the
presence of an external electric or magnetic field may have in this respect.

The energy of a relativistic particle (hadron) with mass $m$ and charge $q$
in a magnetic field aligned in $z$-direction is to leading order given
by (e.g., Ref.~\cite{Luschevskaya:2015cko})
\be
\label{eq:ene-polar}
E^2_{H;n} = m^2 + (1+2n)|qB| - g_H\sz qB - 4\pi m \beta_H |eB|^2 + \ldots \,,
\quad\quad n\in\mathbb{Z}_0^+\,,
\ee
where $g_H$ is the $g$-factor of the hadron $H$ and $\beta_H$ is the polarizability,
which is absent in the free field cases discussed in Sec.~\ref{contiB}. Note
that our convention for the polarizabilities is in agreement with the ones
from~\cite{Luschevskaya:2015cko,Beane:2014ora,Chang:2015qxa},
which use the non-relativistic limit, and it includes a contribution from the tensor
polarizability for a particle with spin~\cite{Beane:2014ora,Chang:2015qxa}.
The ellipses stand for higher-order terms in $|eB|$. The magnetic moment $\mu$
of the particle is related to its $g$-factor in the nonrelativistic limit. The
$g$-factor and the polarizability represent the leading- and next-to-leading-order
responses of the bound state with respect to an external magnetic field. In
particular,
\be
\label{eq:mmom-polar}
g_H = \left. - \frac{1}{\sz} \Big[\frac{\partial E_{H;0}^2}{\partial |eB|} - 1 \Big]
\right|_{B=0} \quad \text{and} \quad \beta_H = \left.
\frac{1}{8\pi m} \frac{\partial^2 E_{H;0}^2}{\partial |eB|^2} \right|_{B=0} \,.
\ee
Here, the derivative $\partial$ is defined as the derivative with respect to $|eB|$
while keeping all other renormalized parameters fixed.
When we measure these with Wilson fermions via the background field method, however,
we obtain the full derivatives, which receive contributions from the implicit
$B$-dependence of the quark mass, so that  
\be
\label{eq:backgr-mmom}
\frac{d E_{H;0}^2}{d |eB|} = \frac{\partial E_{H;0}^2}{\partial |eB|} + \sum_f
\frac{\partial E_{H;0}^2}{\partial m_f} \frac{\partial m_f}{\partial |eB|} =
\frac{\partial E_{H;0}^2}{\partial |eB|} - \sum_f \frac{\partial E_{H;0}^2}{\partial
m_f} \frac{\partial m_{c;f}}{\partial |eB|} \,.
\ee
Similarly, additional terms emerge for the second derivative as well.
Analogous relations hold also for measurements of the electric polarizability.
In Eq.~\refc{eq:backgr-mmom}, the sum is over the quark flavors $f$ present in the
measurement. In the quenched setup the sum includes only contributions from
valence quarks. For dynamical quarks exposed to the external field, however, the
sum also includes effects from sea quarks.

The additional terms including derivatives of the quark masses with respect to the
external field are lattice artifacts and thus vanish in the continuum limit.
They are not present for actions where the quark mass is protected by
a remnant of chiral symmetry. For Wilson fermions, however, those terms are
present and contaminate the results for magnetic moments and
polarizabilities. These contaminations can be removed by tuning the quark mass
along LCP(B)s as discussed above, resulting in a cancellation of the derivatives
of the quark masses with respect to the external field. Even when the tuning is
only approximate, it leads to a strong reduction of the additional terms in
Eq.~\refc{eq:backgr-mmom} and, consequently, a
reduction of lattice artifacts. We would also like to emphasize that this effect
will always be present (and of the same size), no matter how small the external
field is since polarizabilities are derivatives with respect to the field,
Eq.~\refc{eq:mmom-polar}.

\subsection{Magnetic moments and polarizabilities of mesons and the impact of
LCP(B)s}

We proceed by investigating the strength of the
effect mentioned above for polarizabilities and $g$-factors of mesons. Our setup
is not optimal for the extraction of these quantities, since the volume in 
our simulations is rather small and, thus, the
smallest available external field following Eq.~\refc{eq:fquant} is relatively
large. In fact, we could only use the lowest two values of the external field for
the extraction of $g_H$ and $\beta_H$. Obviously, this practice does not lead to a
precision determination of these quantities, but it is sufficient for the purpose
of demonstrating the improvement achieved.

\begin{figure}[t]
\begin{minipage}[c]{.47\textwidth}
 \includegraphics[]{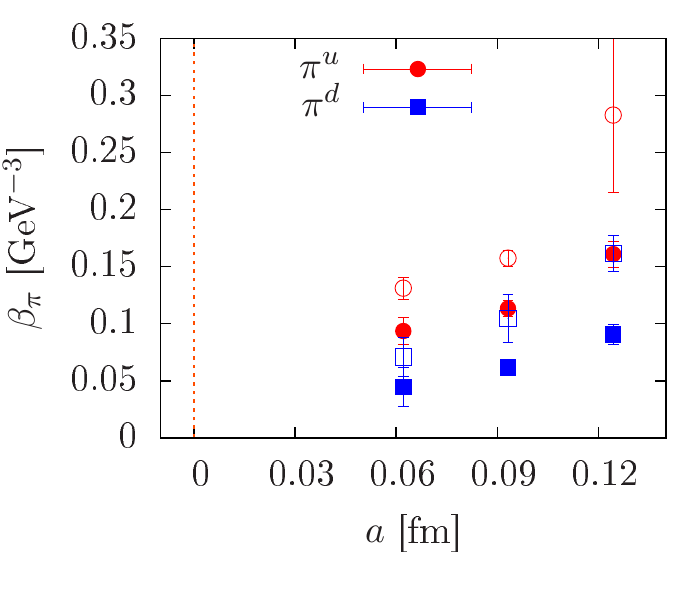}
\end{minipage}
\begin{minipage}[c]{.47\textwidth}
 \includegraphics[]{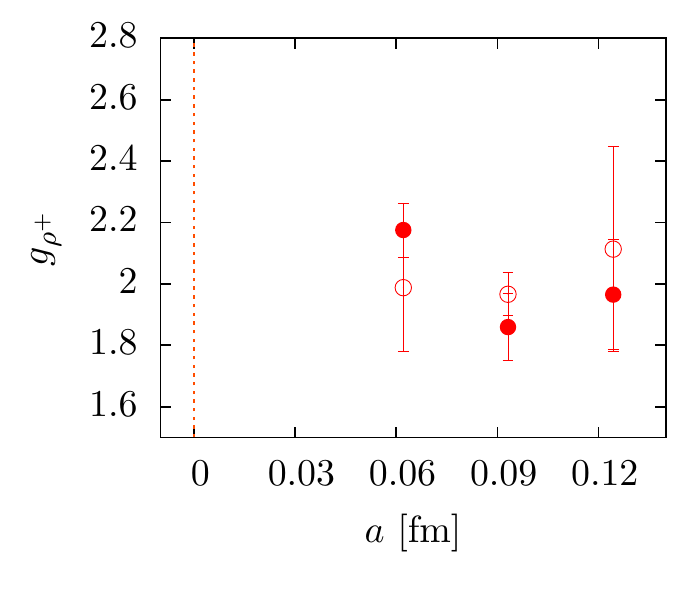}
\end{minipage}
\caption{Results for the polarizabilities of the neutral pions with $\bar{u}u$
and $\bar{d}d$ flavor content and for the magnetic moment of the (charged)
$\rho$-meson versus the lattice spacing. The filled symbols correspond to the results
obtained from LCP(B)s and the open symbols are the results from measurements with
a constant $\kappa$ value.}
\label{fig:polar-a}
\end{figure}

Let us start with the polarizability of the neutral and the charged pion.
Concerning the neutral pion, we can only determine the polarizability of
pions with $\bar{u}u$ and $\bar{d}d$ flavor content individually. The results
versus the lattice spacing are shown in Fig.~\ref{fig:polar-a} (left), both
the ones obtained from LCP(B)s (filled symbols) and the ones obtained by keeping
$\kappa$ constant (open symbols). At finite lattice spacing, there is a visible
difference between the two sets of results, and lattice artifacts appear to be
larger for the results from constant $\kappa$, as expected. We did not
attempt a continuum extrapolation, but it is clear that a naive linear
continuum extrapolation leads to different results for the improved [LCP(B)]
case and the unimproved one. However, the data indicate that a linear continuum
extrapolation is not valid in the regions of lattice spacings at our
disposal. A careful extrapolation to the continuum should, eventually, lead
to the same result for both cases. Comparing our results from LCP(B)s to those
of Ref.~\cite{Luschevskaya:2015cko}, we see that they are in the same ballpark,
in particular, those for the pion with $dd$ flavor content. Concerning the
charged pion, the rather large uncertainties for the masses at finite $B$
lead to large uncertainties for the polarizability, so we cannot draw
any conclusions from them and we will not discuss these results here.

Moving on to the $\rho$-mesons, the polarizability for the $\rho^{u/d}$ with
$\sz=\pm1$ and the $\rho^\pm_0$ can be extracted analogously to the ones for
the neutral and the charged pions, respectively.
However, once more the uncertainties are too large to draw any definite conclusion,
so we exclude these results from the discussion as well.
The situation is somewhat better for the $g$-factor of $\rho^+$ with
$\sz=\pm1$. In this case the energies in Eq.~\refc{eq:ene-polar} allow for
the separation of the effects from the $g$-factor and the polarizability
by building the combinations
\be
\Delta = \frac{1}{2} \big( E^2_{\rho^+_-} - E^2_{\rho^+_+}
\big) \quad \text{and} \quad \Sigma = \frac{1}{2}
\big( E^2_{\rho^+_-} + E^2_{\rho^+_+} \big) \,.
\ee
Following Eq.~\refc{eq:ene-polar}, $\Delta$ is proportional to $g$, while
$\Sigma$ only contains the polarizability $\beta_{\rho^+}$ as a free
parameter up to $O(|eB|^2)$. Note that in the quantity $\Delta$-effects
from the additive quark mass renormalization will cancel, since they appear
with equivalent prefactors in $E^2_{\rho^+_-}$ and
$E^2_{\rho^+_+}$. Consequently, we expect to obtain similar results
for $g_{\rho^+}$ from LCP(B)s and with constant $\kappa$. The results are shown
in Fig.~\ref{fig:polar-a} (right). Indeed, the two sets of results show
good agreement within uncertainties. The results indicate that $g_{\rho^+}$
is close to the free-case value $g=2$, which is in qualitative agreement
with previous lattice
computations~\cite{Hedditch:2007ex,Gurtler:2008zz,Luschevskaya:2016epp}
and findings in chiral perturbation theory~\cite{Djukanovic:2013mka}.

\begin{figure}[t]
\begin{center}
 \includegraphics[]{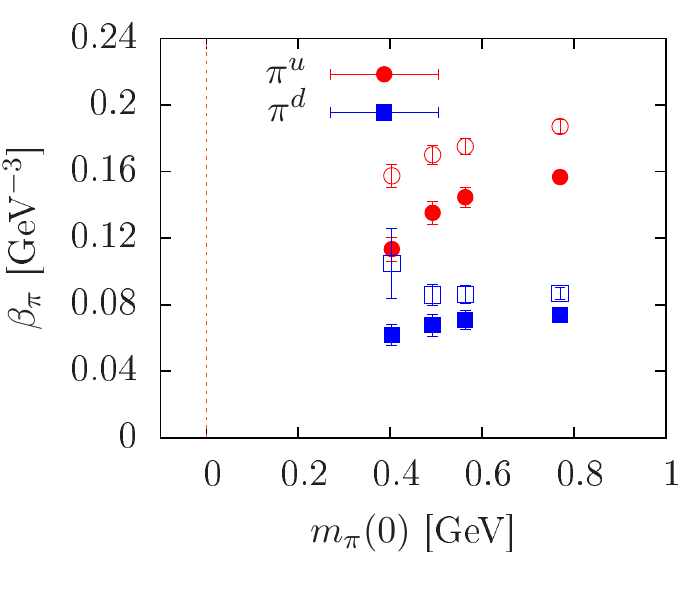}
\end{center}
\vspace*{-.3cm}
\caption{Results for the polarizabilities of the neutral pions with $uu$ and
$dd$ flavor content versus $m_\pi$ at $B=0$. The filled symbols correspond
to the results obtained from LCP(B)s and the open symbols are the results
from measurements with a constant $\kappa$ value. All results were obtained 
on the same lattice ensemble with $\beta=6.0$.}
\label{fig:polar-m}
\end{figure}

Since the relative size of the change in quark mass depends on the bare
quark mass we have at $B=0$ (the change in $m_{c;f}$ is independent of
$\mb_f$), we expect the impact of the additional lattice artifacts to
become increasingly important for smaller quark masses. We
test this intuition by looking at the polarizabilities of the neutral
pions for different values of $m_\pi(B=0)$ on the lattice with
$\beta=6.0$. The results are shown in Fig.~\ref{fig:polar-m}. The
increasing impact when going to smaller quark masses is clearly visible
in the data.
We note that lattice artifacts of this type could potentially also be responsible for the
pion mass dependence of the electrical polarizability of neutral pions obtained 
using the connected part of the correlation function 
only~\cite{Detmold:2009dx,Lujan:2014kia,Lujan:2016ffj}. It has been found that the
polarizability becomes negative for physical quark masses, which disagrees with expectations
from chiral perturbation theory~\cite{Bijnens:1987dc,Donoghue:1988eea,Hu:2007ts}.

\section{Conclusions}
\label{sec:concl}

In this paper, we investigated the meson spectrum in QCD at zero temperature in 
the presence of background magnetic fields $B$
with Wilson fermions in the quenched approximation. The new methods introduced in this paper 
allow us for the first time to perform the continuum limit for the Wilson spectrum at a finite value
of the external field. The novelty of our approach is the introduction of a magnetic 
field-dependent improvement term that ensures that the {\it renormalized} quark mass 
remains independent of $B$. This requirement defines a magnetic field-dependent 
line of constant physics LCP(B) for the {\it bare} quark mass parameter. 
This $B$-dependent tuning of the bare quark mass is absent in the continuum 
(as we discuss in Appendix~\ref{app3}) and only concerns fermion discretizations that suffer
from an additive renormalization, such as Wilson quarks, where this tuning is beneficial already 
in the free case (see Appendix~\ref{app1}). Note that a similar tuning along LCP(B)s should
also be carried out for nonuniform external fields; see the discussion in Appendix~\ref{app1}.
The LCP(B) was determined using the lattice Ward-Takahashi identities for nonvanishing
electromagnetic fields, which we derived in Appendix~\ref{app2}. 

We emphasize that the improvement only differs from the naive approach by lattice 
artifacts, which, however, may be large. 
In particular, we demonstrated that without the improvement meson masses suffer from enormous 
discretization effects for strong magnetic fields.
Besides the impact for the strong field limit, we also considered the effect of the 
improvement for derivatives with respect to the magnetic field at $B=0$ -- i.e.,
magnetic polarizabilities. Also, here we found that our approach suppresses lattice 
artifacts and enables a flatter, controlled continuum extrapolation.

Our most important results about the spectrum 
involve the mass of the neutral (connected) pion, which 
was found to decrease monotonously as the magnetic field grows, and the mass of the 
lightest charged $\rho$-meson, which remains nonzero for the whole range of magnetic fields 
that we considered.
The latter might be relevant for the $\rho$-meson condensation predicted to set in 
for strong magnetic fields~\cite{Schramm:1991ex,Chernodub:2010qx}. 
Nevertheless, we mention again that our quenched analysis of the 
$\rho$-meson mass in constant background magnetic fields cannot capture the subtle details 
of the superconducting vacuum~\cite{Chernodub:2011mc,Chernodub:2013uja}.
Indeed, there are examples in the literature which show that the interplay between
sea and valence quarks can lead to unexpected effects~\cite{Bruckmann:2013oba}.
Eventually, the study of the meson spectrum should thus be repeated including
sea quarks. Note, however, that this demands the generation of new
configurations for each value of the quark mass and the external field in the
(minimal) $N_f=1+1$ setup at large values of $N_t$,
rendering the study with dynamical fermions at least 2 orders of magnitude more
expensive than the present quenched study.
In addition to the investigation of lattice artifacts and quark mass effects, 
we have also checked for finite size effects and
found these to be negligible compared to the other uncertainties (see Sec.~\ref{fse}).

\begin{figure}[t]
 \centering
 \includegraphics[]{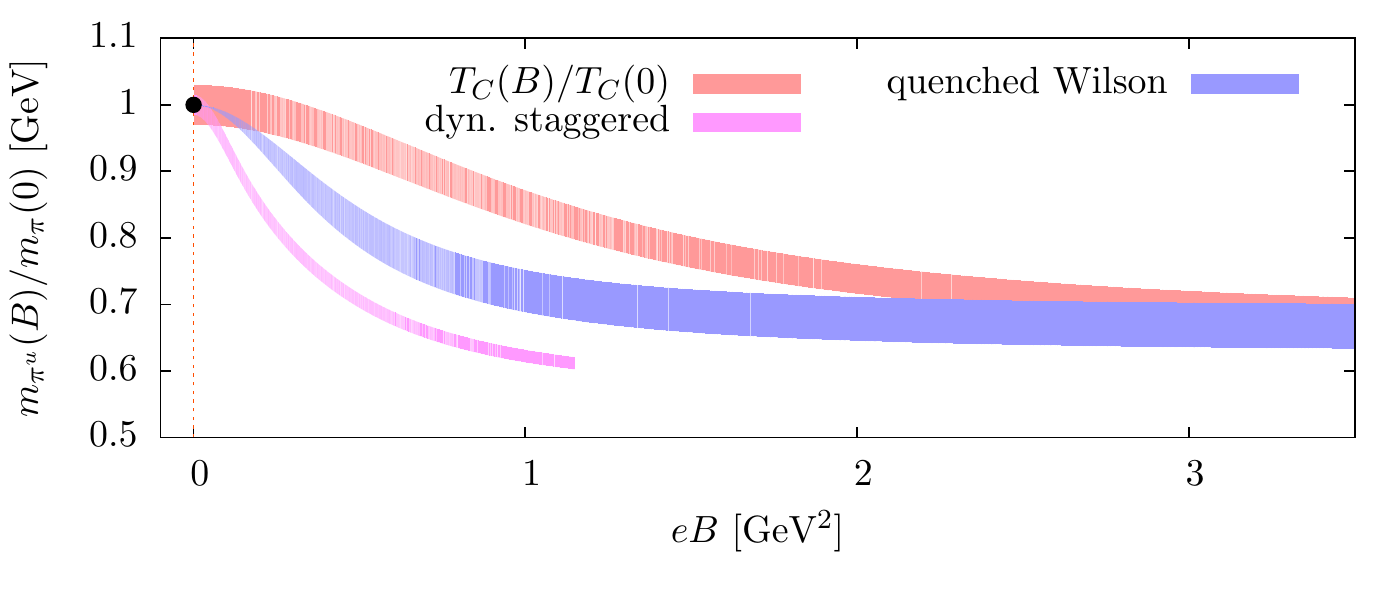}
 \caption{\label{fig:tccomp}Comparison of Wilson and staggered
 continuum extrapolated results for the
 lightest pion mass (normalized by its $B=0$ value). Also shown is
 the QCD transition temperature
 defined via the inflection point of the average light quark condensate --
 also in units of its $B=0$ value.
 Note the difference in the $B=0$ pion mass, which is $415\textmd{ MeV}$
 for the Wilson results and $135\textmd{ MeV}$ for the staggered data.}
\end{figure}

Finally, we elaborate on the connection between the $B$-dependence of the
mass of the lightest hadron and that of the QCD transition temperature.
The latter was determined using continuum extrapolated dynamical staggered
quarks with physical masses in Refs.~\cite{Bali:2011qj,Endrodi:2015oba}.
Remember that according to our results the quantity $m_{\pi^u}(B)/m_{\pi}(0)$
has only mild quark mass dependence, cf.\ Fig.~\ref{fig:pions-qmdep}.
It is thus sensible to compare our continuum extrapolated results to
the staggered results using physical pion masses.
Employing the same definition involving connected neutral pion correlators,
the Wilson and staggered results\footnote{The details of the staggered
simulation setup are described in Refs.~\cite{Borsanyi:2010cj,Bali:2011qj}. The
determination of $m_{\pi^u}$ follows the strategy of Ref.~\cite{Bali:2011qj} for the
measurement of $m_{\pi^+}$ and is based on the same lattice ensembles.} for
$m_{\pi^u}(B)/m_{\pi}(0)$ are shown in Fig.~\ref{fig:tccomp}.
For $eB \lesssim 1$~GeV$^2$, where both results are available,
the comparison reveals small differences due to the different
$B=0$ pion masses and the inclusion of sea effects in the staggered case.

Regarding the QCD transition temperature, we considered the parameterization
of the crossover transition temperature $T_c(B)$ defined using the
inflection point of the average light quark condensate~\cite{Endrodi:2015oba}.
According to Fig.~\ref{fig:tccomp}, the $B$-dependence of the curves looks
qualitatively very similar -- with an initial reduction followed by a
saturation to around $60\%$ to $70\%$ of the $B=0$ value. This supports
the picture sketched in the Introduction, where we compared the finite
temperature QCD transition to the ``melting'' of the lightest hadron state.
Note that the quantitative difference for intermediate values of $B$ could
originate from effects due to the charged pions, which, in this region, are
potentially light enough to have a direct influence on the transition
temperature. We remark that, while the inclusion of charged sea quarks appears to have
a marginal impact on the $T=0$ pion masses, it is known to make a drastic
difference for $T_c$~\cite{Andersen:2014xxa}.
A possible explanation is that the transition temperature is
driven by bulk effects encoded in the configurations (i.e., the action used to
generate them), so an action leading to the correct pion mass is essential
to observe the correct features of the transition. For meson masses, measurements
are most affected by the properties of valence quarks, which reflects itself
in the fact that the quenched spectrum reproduces the QCD spectrum up to about
10\% accuracy (see, e.g., Refs.~\cite{Aoki:1999yr,Aoki:2002uc}). Nevertheless, it would
be interesting to understand this subtle difference between measurements of
properties of the phase transition and the hadron spectrum in more detail.

\acknowledgments

This research was funded by the DFG (Emmy Noether Programme EN 1064/2-
1 and SFB/TRR 55). The majority of the simulations was performed on the iDataCool 
cluster of the Institute for Theoretical Physics at the University of Regensburg.
The authors are grateful for the useful correspondence with Pavel Buividovich,
Maxim Chernodub, Davide Giusti, Rainer Sommer, and Arata Yamamoto. We thank Max
Theilig for a careful reading of the Appendixes.

\appendix

\section{Additive mass renormalization in the free case}
\label{app1}

To demonstrate the presence of a $B$-dependent additive quark mass renormalization,
it is instructive to look at the free case.
We have already discussed this setting in Ref.~\cite{Bali:2015vua}; here, we repeat the 
main findings for the sake of completeness.

\begin{figure}[t]
 \centering
 \includegraphics[width=7cm]{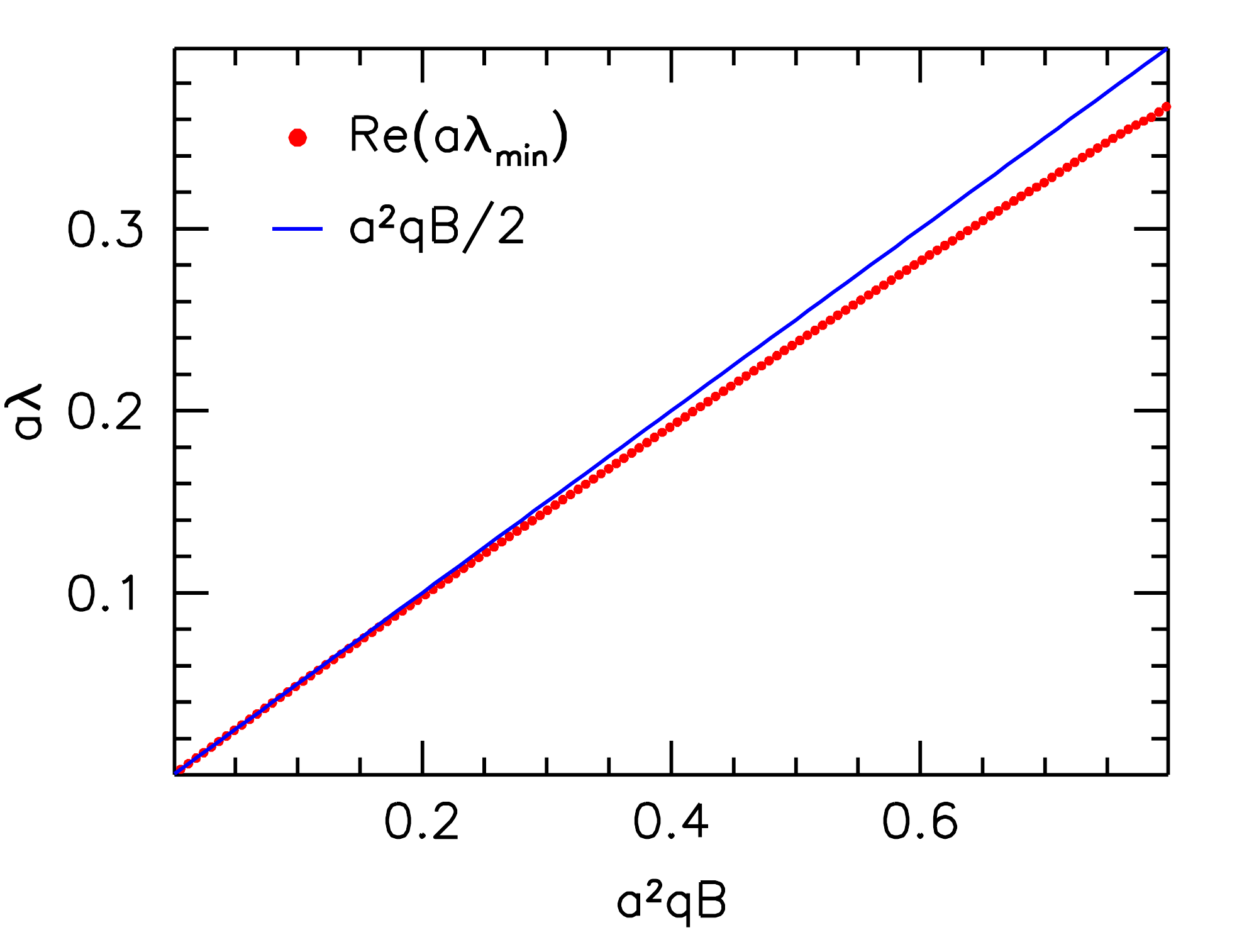}
 \caption{\label{fig:wilsonmr}The lowest eigenvalue (in lattice units) of the 
 Wilson Dirac operator with $r=1$ against the magnetic field (also in lattice units),
 calculated on a two-dimensional $32^2$ lattice.}
\end{figure}

The massless Wilson Dirac operator may be written schematically as $D_W =
\slashed{D} + ar\cdot \Delta/2$,  where $\slashed{D}$ is the naive discretization of 
the (anti-Hermitian) continuum Dirac operator and
the second part is the Wilson term. While $\slashed{D}$ describes spin-$1/2$ particles,
$\Delta$ is the discretization of the Klein-Gordon operator and, thus, describes
scalars. The eigenvalues of the two operators for $B>0$ can be read off from
Eqs.~(\ref{eq:Efermion}) and~(\ref{eq:Escalar}) for massless particles.
Therefore (at zero temperature, where the lowest Matsubara frequency is zero),
the lowest eigenvalue of $\slashed{D}$
vanishes, while that of the Wilson term 
equals $ar\cdot|qB|/2$. Since the two operators commute on the subspace spanned
by the lowest eigenmode, the real part of the lowest eigenvalue 
of $D_W$ increases linearly with $B$. When discretized on the lattice, this 
conclusion continues to hold for small values of $a^2qB$; 
see Fig.~\ref{fig:wilsonmr}. In fact, for this check, 
it suffices to diagonalize the Wilson operator on a two-dimensional $x,y$-plane, since
the eigenmodes factorize and form plane waves in the $z$- and $t$-directions.

This increase of the lowest eigenvalue is
equivalent to an additive shift in the quark mass, implying a $B$-dependent
additive renormalization. The critical mass (or, the critical hopping parameter
$\kappa_c$) is reduced by the magnetic field 
as
\begin{equation}
am_{c;f}(a,B) = am_{c}(a) -  a^2|q_fB|/2 \,, \qquad \kappa_{c;f}^{-1}(a,B) =
\kappa_c^{-1}(a) - a^2 |q_fB| \,.
\label{eq:free_tuning}
\end{equation}
Thus, in the free theory, a fixed renormalized quark mass, see
Eq.~(\ref{eq:massreno}), is achieved if the quark mass parameter is shifted by
the same amount (note that in the free case $Z_m=1$).
We mention that this additive shift in the spectrum is only present for
fermion formulations which break the full chiral symmetry explicitly and does
not appear, for example, in the staggered formulation~\cite{Endrodi:2014vza}.

The effect of this $B$-dependent tuning can be demonstrated using the 
free ``pion masses,'' i.e., the energies associated to the leading
decay of the pseudoscalar correlation functions, corresponding to the energies
of quark-antiquark states with imposed pion quantum numbers. On the one hand,
for neutral correlation functions, both quarks have magnetic moments parallel 
to $\ve B$ and thus are in
the ground state. The associated energy should be $E(B)=m_u+m_d$. 
On the other hand, for the charged correlation function, one
of the quarks (that with the smaller absolute charge) is forced to have 
its magnetic moment antiparallel to $\ve B$. The energy in this case
should be $E(B)=m_u+\sqrt{m_d^2+2|q_dB|}$. 

\begin{figure}[t]
\begin{minipage}[c]{.48\textwidth}
\centering
 \vspace*{-4mm}
\includegraphics[]{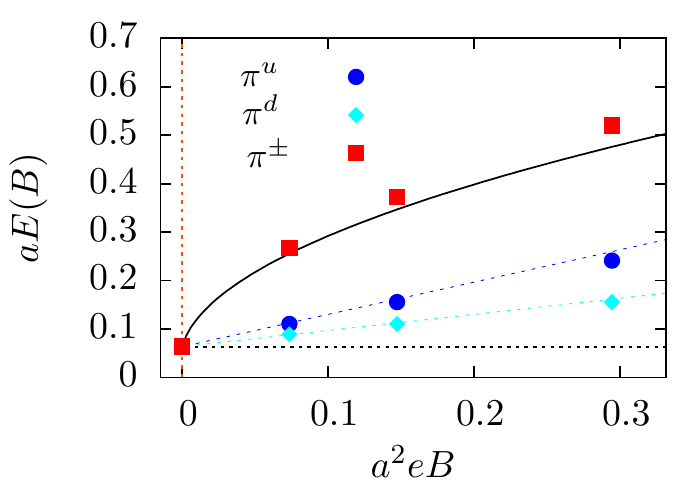}
\end{minipage}
\begin{minipage}[c]{.48\textwidth}
\centering
 \vspace*{-4mm}
\includegraphics[]{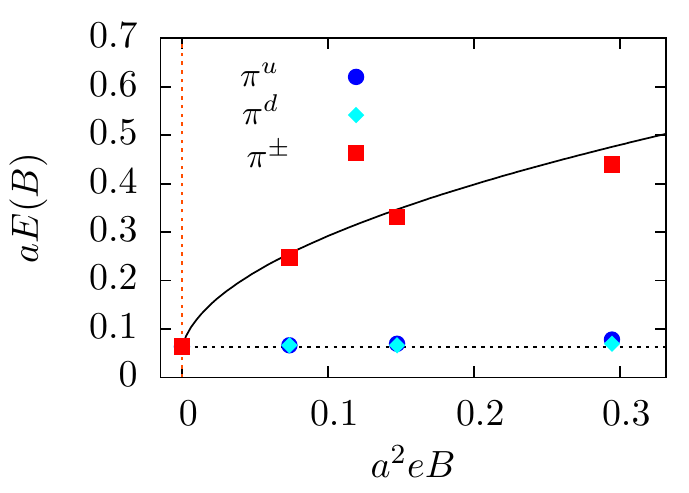}
\end{minipage}
 \vspace*{-1mm}
\caption{Results for the energies associated with free correlation functions in
the pseudoscalar channel without (left) and with (right) tuning of $\kappa$
with $B$ along the trajectory~(\protect\ref{eq:free_tuning}). The colored
dashed lines in the left plot show the analytic expectations
$E_f(B)=E(0)+a|q_fB|/2$ for energies associated with neutral correlation
functions, and the solid line is the expectation for charged pions.}
\label{fig:free-pions}
\end{figure}

Our numerical results for the energies at fixed $\kappa=0.124$ are shown in
Fig.~\ref{fig:free-pions} (left). The energy of the neutral pion increases
with the magnetic field, indicating the unphysical increase of the quark mass by
the amount $a^2|q_fB|/2$. Tuning the hopping parameters along the
trajectory~(\ref{eq:free_tuning}) instead, the neutral pion mass remains
constant as it should; see Fig.~\ref{fig:free-pions} (right). Our results are
also in good agreement with the expectation for the charged ``pion.''

The shift in the bare quark mass basically follows from the magnetic
field dependence of the lowest energy state of a noninteracting charged
scalar particle. For a homogeneous background magnetic field, this
dependence can be found analytically
and is given by Eq.~(\ref{eq:Escalar}). We also checked numerically what
happens if the magnetic background field is inhomogeneous. In particular, we
considered an oscillatory field $B(x)\propto \sin (2\pi k x/L_x)$ with
$k\in\mathbb{Z}$ and a half-half field $B(x)\propto (-1)^{x \textmd{ div }
(L_x/2)}$. In both cases
the bare quark mass was observed to increase with growing magnetic field
-- although not linearly, as for the homogeneous background, but
quadratically in $a^2qB$.

\section{Axial Ward-Takahashi identities for QCD+QED}
\label{app2}

In this Appendix we will derive axial and vector WIs
for QCD in the presence of electromagnetic interactions with two light
quark flavors, i.e., for QCD+QED. We will start by (re)deriving the continuum
WIs in Appendix~\ref{app2.1}, which, to our knowledge, have so far only been
discussed in Ref.~\cite{Scherer:2002tk}, before we discuss the WIs for (unimproved)
Wilson quarks on the lattice in Appendix~\ref{app2.2}. Note that an account of
the continuum WIs has already appeared in Ref.~\cite{Bali:2015vua}. The WIs for
$\Ord(a)$-improved Wilson quarks can be derived along the same lines and will be
discussed briefly in Appendix~\ref{app2.3}. Note that the axial Ward identity
for the domain wall fermion discretization has been derived
in Ref.~\cite{Blum:2007cy} and recently for twisted mass fermions at maximal twist
in Ref.~\cite{Giusti:2017jof}.\footnote{We thank Davide Giusti for drawing
our attention to this.}

\subsection{Continuum Ward-Takahashi identities}
\label{app2.1}

In the continuum, the Euclidean fermion action including $u$ and $d$ quarks in
the presence of QED interactions is given by
\be
\label{eq:conti-act}
S_F^{\rm conti} = \int d^4 x \, \bp(x) (\slashed{D} + M) \p(x)
\ee
with the covariant derivative
\be
\label{eq:conti-cder}
\slashed{D}=\slashed{D}_{QCD} + \slashed{D}_{\rm em} \,, \qquad
\slashed{D}_{\rm QCD} = \gm \big( \partial_\mu + i G_\mu(x) \big) \ve{1} \,,
\qquad \slashed{D}_{\rm em} = i Q \gamma_\mu A_\mu(x)  \,;
\ee
$M$ the matrix of bare quark masses in the continuum, given by
Eq.~\refc{eq:mass-mat}; and the charge matrix
\be
\label{eq:charge-mat}
Q = e \Big( \frac{\tau^0}{6} + \frac{\tau^3}{2} \Big) \quad
\text{with} \quad \tau^0 \equiv \mathbf{1} \,.
\ee
We have denoted the gluon fields by $G_\mu(x)$ and the
electromagnetic (photon) field by $A_\mu(x)$.

Ward-Takahashi identities can be derived by varying the expectation
value of some test operator $O$ under transformations of the fermionic
variables in the path integral of the form
\be
\dep(x) = i \Big[ \vda \frac{\tau^j}{2} \Big] \p(x) \,, \quad \debp(x) = -i \bp(x)
\Big[ \vda \frac{\tau^j}{2} \Big] \,,
\ee
corresponding to infinitesimal transformations under $\SU_V(2)$, and 
\be
\label{eq:axial-trafo}
\dep(x) = i \big[ \ada \frac{\tau^j}{2} \gf \big] \p(x) \,, \quad \debp(x) = i
\bp(x) \big[ \ada \frac{\tau^j}{2} \gf \big]
\ee
for infinitesimal transformations under $\SU_A(2)$. Here, $\tau^j$ is any of the
three Pauli matrices for $j=1,2,3$ and the unit matrix for $j=0$, corresponding
to isosinglet transformations. Using these transformations one obtains (with $X=A,V$)
\be
\label{eq:expval-wi}
\frac{\partial}{\partial \alpha^j_X(x)} \expv{O} = \expv{\frac{\partial\,O}
{\partial \alpha^j_X(x)}} - \expv{\frac{\partial\,S_F}
{\partial \alpha^j_X(x)}\,O} = - \expv{\frac{\partial\,S_F}
{\partial \alpha^j_X(x)}\,O} = 0 \,,
\ee
in the flavor nonsinglet case, where the second equality is valid as long as
the operator $O$ has no support at the point $x$. In the singlet case, one also has to consider
the variation of the measure.

Using Eqs.~\refc{eq:conti-act} and~\refc{eq:conti-cder}, we obtain
\be
\label{eq:conti-variat}
\frac{\partial\,S^{\rm conti}_F}{\partial \alpha^j_X(x)} = \int d^4x \, \frac{\partial}
{\partial \alpha^j_X(x)} \Big( \bp(x) (\slashed{D}_{\rm QCD} +M) \p(x) \Big) + \int d^4x \, 
\frac{\partial} {\partial \alpha^j_X(x)} \Big( \bp(x) \slashed{D}_{\rm em}
\p(x) \Big)
\ee
The result for the first term is well known for the two types of transformations
(see, e.g., Ref.~\cite{Vladikas:2011bp}),
\begin{eqnarray}
\label{eq:conti-variatV}
\frac{\partial} {\partial \alpha^j_V(x)} \Big( \bp(x) (\slashed{D}_{\rm QCD} +
M) \p(x) \Big) & = & - i \partial_\mu (J_V)_\mu^j(x) + i \bp(x) \com{M}{\tau^j/2} \p(x)
\quad \text{and} \\
\label{eq:conti-variatA}
\frac{\partial} {\partial \alpha^j_A(x)} \Big( \bp(x) (\slashed{D}_{\rm QCD} +
M) \p(x) \Big) & = & - i \partial_\mu (J_A)^j_\mu(x) + i \bp(x) \acom{M}{\tau^j/2}
\gf \p(x) \,,
\end{eqnarray}
where the curly brackets denote the anticommutator, so we only need to
compute the additional terms stemming from the second term in
Eq.~\refc{eq:conti-variat}. Here, $(J_V)_\mu^j(x)$ is the (continuum) vector current
from Eq.~\refc{eq:vect-current}, including the Pauli matrix $\tau^j/2$,
and $(J_A)^j_\mu(x)$ is the local axial vector current\footnote{We use
the notation $J_V$ and $J_A$ instead of $V$ and $A$ in this section to avoid
confusion with the electromagnetic vector potential $A_\mu$ and to distinguish
between point-split and local currents in the lattice regularization.}
\be
(J_A)^j_\mu(x) = \bar{\psi}(x) \gm \gf \frac{\tau^j}{2} \psi(x) \,.
\ee

Considering vector transformations, the additional terms are given by
\be
\frac{\partial} {\partial \alpha^j_V(x)} \Big( \bp(x) \slashed{D}_{\rm em}
\p(x) \Big) = - \frac{e}{4} \bp(x) \gm A_\mu(x) \com{\tau^3}{\tau^j} \p(x)
= - i \etjk e \bp(x) A_\mu(x) \gm \frac{\tau^k}{2} \p(x) \,,
\ee
so together with
\be
\com{M}{\tau^j/2} = \frac{1}{4} (m_u-m_d) \com{\tau^3}{\tau^j} = i \etjk
(m_u-m_d) \frac{\tau^k}{2}
\ee
we arrive at the continuum vector WI
\be
\label{eq:VWI-conti}
\partial_\mu (J_V)^j_\mu(x) = i\etjk (m_u-m_d) S^k(x)
- \etjk e A_\mu(x) (J_V)^k_\mu(x) \,.
\ee
Here, $\epsilon_{ijk}$ is the totally antisymmetric tensor for $i,j,k=1,2,3$
and zero if either of $i,j,k=0$, and we have introduced the scalar density
\be
\label{eq:scalarD}
S^j(x) = \bp(x) \frac{\tau^j}{2} \p(x) \,.
\ee

With axial vector transformations, we obtain
\be
\frac{\partial} {\partial \alpha^j_A(x)} \Big( \bp(x) \slashed{D}_{\rm em}
\p(x) \Big) = - \frac{e}{4} \bp(x) A_\mu(x) \acom{\gm\tau^3}{\gf\tau^j}
\p(x) = - i \etjk e \bp(x) A_\mu(x) \gm \gf \frac{\tau^k}{2} \p(x)
\ee
and
\be
\acom{M}{\tau^j/2} = (m_u+m_d) \frac{\tau^j}{2} + \frac{1}{2} (m_u-m_d) \delta_{j3} \,,
\ee
so we get for the continuum axial WI
\be
\label{eq:AWI-conti}
\begin{array}{rl} \displaystyle \partial_\mu (J_A)^j_\mu(x) = & \displaystyle
 (m_u+m_d) P^j(x) + \delta_{j3} (m_u-m_d) P^0(x) - \etjk e A_\mu(x)
 (J_A)^k_\mu(x) \\
 - & \displaystyle \frac{e^2}{16\pi^2} \varepsilon^{\alpha\beta\mu\nu}
 F_{\alpha\beta}(x)  F_{\mu\nu}(x) \text{Tr}\big[\tau^jQ^2\big]
 - \delta_{j0} \frac{1}{16\pi^2} \varepsilon^{\alpha\beta\mu\nu}
 \text{Tr}\big[H_{\alpha\beta}(x)  H_{\mu\nu}(x)\big] \,.
 \end{array}
\ee
Here we have introduced the pseudoscalar densities
\be
\label{eq:pseudoD}
P^j(x) = \bp(x) \gf \frac{\tau^j}{2} \p(x) \quad \text{and} \quad P^0(x) =
\bp(x) \gf \frac{\ve{1}}{2} \p(x) \,;
\ee
$H_{\mu\nu}$ and $F_{\mu\nu}$ are the gluonic and electromagnetic field strength
tensors, respectively; and Tr denotes the trace over flavor and color indices.
The terms in the last line of Eq.~\refc{eq:AWI-conti}
are the ones associated with the Jacobian of the transformation
from Eq.~\refc{eq:axial-trafo}~\cite{peskin1995introduction},
known as the axial anomaly. For QCD+QED, there are two such terms, associated with
the topological charge operators for the gluonic and the electromagnetic fields,
respectively. Note that for external magnetic fields (and no electric fields), the
electromagnetic anomaly is absent.

\subsection{Ward-Takahashi identities for Wilson fermions}
\label{app2.2}

We will now derive similar identities for (unimproved) Wilson fermions using
the fermion action from Eq.~\refc{eq:Wilson-act}. In particular, we need to
compute $(\partial\,S_F)/(\partial \alpha^j_X(x))$ for the transformations from
Eqs.~\refc{eq:conti-variatV} and~\refc{eq:conti-variatA}. The basic idea to
separate the resulting equations in parts which are already known and new parts
is the same as in the continuum case. In practice, however, the separation is
a bit less intuitive since the electromagnetic field enters the action via the
electromagnetic link variables $u_\mu(x)$; see Eq.~\refc{eq:u3-links}. The
separation can be done by bringing the Pauli matrices from the variation of
the $\bp$ variables to the right-hand side of the link
variables.\footnote{Note that this choice corresponds to a particular
convention for the definition of point-split bilinear operators. An
alternative convention includes the Pauli matrices to the left of the link
variables. Both definitions lead to equivalent results as long as one strictly
follows one and the same convention analytically and numerically.}
When the links $u_\mu(x)$ are absent, this is possible since the
gluonic link variables commute with matrices in flavor space. In the presence
of $u_\mu(x)$, new terms will appear due to the nonvanishing commutator
\be
\label{eq:utau-commut}
\com{\uam}{\tau^j} = i \etjk \Big[ \exp\Big(i\frac{2}{3}eaA_\mu(x)\Big) - 
\exp\Big(-i\frac{1}{3}eaA_\mu(x)\Big) \Big] \tau^k \equiv i \etjk \, \Duam(x)
\, \tau^k \,,
\ee
where the last equation defines $\Duam(x)$.

Let us start again with the vector transformations. Using the results for
Wilson fermions from Ref.~\cite{Karsten:1980wd}, we arrive at the vector WI
\be
\label{eq:VWI-lattice}
\begin{array}{rl}
\displaystyle a^4 \sum_\mu \nabla_x^\mu (\widetilde{J}_V)_\mu^j(x) = &
\displaystyle i \etjk a^4 (m^0_u-m^0_d) S^k(x) \\
 & \displaystyle \left. \begin{array}{rl}
   - i \,r\, \etjk \frac{a^3}{2} \sum_\mu & \Big[ \bp(x)
   \um(x) \Duam(x) \frac{\tau^k}{2} \p(x+\hmu) \vspace*{2mm} \\
   & \;\; + \bp(x) \dum(x-\hmu) \Dduam(x-\hmu)
   \frac{\tau^k}{2} \p(x-\hmu) \Big]
   \end{array} \right\} (1) \vspace*{2mm} \\
 & \displaystyle \left. \begin{array}{rl}
   - i \etjk \frac{a^3}{2} \sum_\mu & \Big[ - \bp(x)
   \gm \um(x) \Duam(x) \frac{\tau^k}{2} \p(x+\hmu) \vspace*{2mm} \\
   & \;\; + \bp(x) \gm \dum(x-\hmu) \Dduam(x-\hmu)
   \frac{\tau^k}{2} \p(x-\hmu) \Big]
   \end{array} \right\} \,. (2)
\end{array}
\ee
In Eq.~\refc{eq:VWI-lattice}, we have introduced the asymmetric lattice
derivative $a\nabla_x^\mu f(x)=f(x)-f(x-\hmu)$, and $\widetilde{J}_V$ is
the point-split (conserved) vector current
\be
\label{eq:psV}
(\widetilde{J}_V)_\mu^j(x) = \frac{1}{2} \Big[ \bp(x) (\gm-r) U_\mu(x)
\frac{\tau^j}{2} \p(x+\hmu) + \bp(x+\hmu) (\gm +r) U^\dagger_\mu(x)
\frac{\tau^j}{2} \p(x) \Big] .
\ee
The first two terms are the standard terms appearing in the vector
Ward identity, while the terms denoted by $(1)$ and $(2)$ appear due to
Eq.~\refc{eq:utau-commut}.

Similarly we obtain for the axial WI
\be
\label{eq:AWI-lattice}
\begin{array}{rl}
\displaystyle a^4 \sum_\mu \nabla_x^\mu (\widetilde{J}_A)_\mu^j(x) = &
\displaystyle a^4 (m^0_u+m^0_d) P^j(x) + \delta_{j3} (m^0_u-m^0_d) P^0(x)
+ a^4 \,r\, X^j(x) \\
 & \displaystyle \left. \begin{array}{rl}
   - i \,r\, \etjk \frac{a^3}{2} \sum_\mu & \Big[ \bp(x)
   \um(x) \Duam(x) \gf \frac{\tau^k}{2} \p(x+\hmu) \vspace*{2mm} \\
   & \;\; + \bp(x) \dum(x-\hmu) \Dduam(x-\hmu) \gf
     \frac{\tau^k}{2} \p(x-\hmu) \Big]
   \end{array} \right\} (1) \vspace*{2mm} \\
 & \displaystyle \left. \begin{array}{rl}
   - i \etjk \frac{a^3}{2} \sum_\mu & \Big[ - \bp(x)
   \gf \gm \um(x) \Duam(x) \frac{\tau^k}{2} \p(x+\hmu) \vspace*{2mm} \\
   & \;\; + \bp(x) \gf \gm \dum(x-\hmu) \Dduam(x-\hmu)
   \frac{\tau^k}{2} \p(x-\hmu) \Big] ,
   \end{array} \right\} (2)
\end{array}
\ee
where
\be
X^j = - \frac{1}{2a} \sum_\mu \Big\{ \Big[ \bp(x) U_\mu(x) \gf \frac{\tau^j}{2}
\p(x+\hmu) + \bp(x+\hmu) U_\mu(x) \gf \frac{\tau^j}{2} \p(x) \Big] + \big[ x\to
x-\hmu \big] - 4 \bp(x) \gf \frac{\tau^j}{2} \p(x) \Big\}
\ee
is the term associated with the variation of the Wilson term
in the standard QCD axial WI (see Ref.~\cite{Bochicchio:1985xa}) and
$\widetilde{J}_A$ is the point-split axial vector current
\be
\label{eq:psA}
(\widetilde{J}_A)_\mu^j(x) = \frac{1}{2} \Big[ \bp(x) \gm \gf U_\mu(x)
\frac{\tau^j}{2} \p(x+\hmu) + \bp(x+\hmu) \gm \gf U^\dagger_\mu(x)
\frac{\tau^j}{2} \p(x) \Big] .
\ee
Note, that in QCD [without terms $(1)$ and $(2)$], the presence of $X^j$ is
responsible for the additive quark mass renormalization since $X^j$ mixes
with the pseudoscalar density. In Eq.~\refc{eq:AWI-lattice}, we have neglected
the terms associated with the anomaly [see Eq.~\refc{eq:AWI-conti}].

Let us briefly discuss the properties and the nature of the terms
denoted by $(1)$ and $(2)$. These two types of terms are due to the
presence of the electromagnetic interactions in the QCD+QED Lagrangian and
vanish, as expected, in the pure QCD case, i.e., when $A_\mu(x)=0$
[since $\Duam(x)=0$]. Since the additional terms are proportional to $\etjk$,
they only contribute to ``charged'' WIs, i.e., WIs which include combinations
of $\tau^1$ and $\tau^2$. Neutral WIs, i.e., such that are a combination
of WIs with $\tau^3$ and $\ve{1}$, are unaffected by the presence of
the QED interactions.\footnote{This is true up to the anomalous terms
for the isosinglet axial WI.} The terms denoted by $(1)$ arise from the Wilson
term (consequently, they are multiplied by a factor of $r$), and, as $X^j$,
they are proportional to a dimension-5 operator and vanish in the
continuum. However, as $X^j$, these terms also mix with the scalar (for the
vector WI) and pseudoscalar (for the scalar WI) density, respectively,
and thus lead to an additional additive quark mass renormalization.
In particular, the presence of such a
term in the vector WI indicates that the quark mass difference does not
vanish when the bare quark masses coincide, meaning that $u$ and
$d$ quark masses renormalizes with different additive terms.
The terms denoted by $(2)$ are operators of dimension 4 and survive the
continuum limit. In fact, they exactly resemble the additional terms
in the continuum WIs, Eqs.~\refc{eq:VWI-conti} and~\refc{eq:AWI-conti},
proportional to $A_\mu(x)$.

To compute current quark masses,
terms of type $(2)$ should be included explicitly in their definition,
since these terms are also present in the continuum WI, while terms of
type $(1)$ should be left out (similar to $X^j$). For an example, consider
the definition of the sum of $u$ and $d$ current quark masses via the
axial WI evaluated for the matrix $\tau^+$ from
Sec.~\ref{theory-masses}. Using the pseudoscalar density $P^+$ from
Eq.~\refc{eq:pseudoD} as the operator $O$ in Eq.~\refc{eq:expval-wi},
a suitable definition for the sum of current quark masses is given by
\be
\label{eq:charged-awi}
a(\mpc_u(B)+\mpc_d(B)) = \frac{ \partial_0
\Big\langle (\widetilde{J}_A)_0^+ (x_0) P^+(0) \Big\rangle -
\Big\langle \,(2)\, P^+(0) \Big\rangle }{
\expv{P^+(x_0) P^+(0)} } \,,
\ee
where $(2)$ refers to the insertion of the associated operators from
Eq.~\refc{eq:AWI-lattice}.

Note that the electromagnetic link variables that need to be included
in the point-split currents depend on the particular flavor matrix for
which the WI is evaluated. As an example, let us once more consider
the WIs for the matrix $\tau^+$. Writing
\be
\uam(x) = \left( \begin{array}{cc} u_\mu^u(x) & 0 \\ 0 & u_\mu^d(x)
\end{array} \right) \quad \text{with} \quad 
u_\mu^f=\exp\big(iq_faA_\mu(x)\big) \,,
\ee
we obtain
\be
\bp(x) u_\mu(x) \tau^+ \p(x) = \bar{d}(x) u_\mu^d(x) u(x) \,.
\ee

\subsection{Ward-Takahashi identities for O(a)-improved Wilson fermions}
\label{app2.3}

For $O(a)$-improvement in QCD+QED, we need to include two
Sheikholeslami-Wohlert~\cite{Sheikholeslami:1985ij} terms in the action,
\be
S_F^{SW} = -\frac{i}{2} \csw a^5 \bp(x) \sigma_{\mu\nu} H_{\mu\nu}(x)
\p(x) \quad \text{and} \quad S_F^{\rm em} = -\frac{i}{2} \cem a^5 \bp(x)
\sigma_{\mu\nu} C_{\mu\nu}(x) \p(x) \,,
\ee
where $H_{\mu\nu}$ is (a suitable discretization of) the gluonic field
strength tensor and $C_{\mu\nu}=QF_{\mu\nu}$ is the electromagnetic one,
multiplied by the electric charge matrix $Q$. The terms in
the WIs which result from these additional terms depend on the particular
choice of discretization. Most commonly used is the clover discretization
of the field strength tensor,
\be
\label{eq:clover}
H_{\mu\nu}(x) = \frac{1}{8a^2} \big[ Q_{\mu\nu}(x) -
Q_{\nu\mu}(x) \big] \quad \text{with} \quad 
Q_{\mu\nu}(x) = U^G_{\mu,\nu}(x) + U^G_{\nu,-\mu}(x) + U^G_{-\mu,-\nu}(x) +
U^G_{-\nu,\mu}(x) \,,
\ee
and similarly for $C_{\mu\nu}$ with $U^G$ replaced by $u$.
Here, $U^G_{\pm\mu,\pm\nu}(x)$ [and similarly $u_{\pm\mu,\pm\nu}(x)$ for
$C_{\mu\nu}$] denotes the multiplication of links around
a plaquette in the $\pm\mu,\pm\nu$-direction, starting from point $x$.

For vector transformations, we only get a contribution from the
electromagnetic clover term. In particular, we get terms which are
proportional to the commutator
\be
\com{u_{\pm\mu,\pm\nu}(x)}{\tau^j} = i \etjk \Big[
\exp\Big(i\frac{2}{3}eaA_{\pm\mu,\pm\nu}(x)\big) -
\exp\Big(-i\frac{1}{3}eaA_{\pm\mu,\pm\nu}(x)\Big) \Big] \tau^k
\equiv i \etjk (\delta u)_{\pm\mu,\pm\nu}(x) \tau^k \,,
\ee
where $A_{\pm\mu,\pm\nu}(x)$ denotes the sum of the phase factors
appearing in the plaquette $u_{\pm\mu,\pm\nu}(x)$ and the last
equation defines $(\delta u)_{\pm\mu,\pm\nu}(x)$ in analogy to
$\Duam(x)$. When we define $(\delta C)_{\mu\nu}$ similar to $H_{\mu\nu}$
from Eq.~\refc{eq:clover} with $U^G_{\pm\mu,\pm\nu}(x)\to
(\delta u)_{\pm\mu,\pm\nu}(x)$, the additional term on the right-hand
side of Eq.~\refc{eq:VWI-lattice} is given by
\be
a^4 \sum_\mu \nabla_x^\mu (\widetilde{J}_V)_\mu^j(x) = 
\text{\refc{eq:VWI-lattice}} + \frac{1}{2} \etjk \cem a^5
\bp(x) \sigma_{\mu\nu} (\delta C)_{\mu\nu} \frac{\tau^k}{2} \p(x) \,.
\ee

For axial vector transformations, it is the anticommutator of $\tau^j$
and the clover term which is relevant for the WI. For the gluonic
clover term the anticommutator is nonvanishing, leading to the known
additional term
\be
\frac{\partial S_F^{\rm SW}}{\partial\ada} = \csw a^5 \bp(x)
\sigma_{\mu\nu}H_{\mu\nu}\gf\frac{\tau^j}{2} \p(x) \,.
\ee
For the electromagnetic clover term we obtain terms proportional to
\be
\begin{array}{rl}
\displaystyle \acom{u_{\pm\mu,\pm\nu}(x)}{\tau^j} = &
\displaystyle \Big[ \exp\Big(iea\frac{2}{3}A_{\pm\mu,\pm\nu}(x)
\Big) + \exp\Big( -iea\frac{1}{3}A_{\pm\mu,\pm\nu}(x) \Big) \Big]
\tau^j + \delta_{3j} (\delta u)_{\pm\mu,\pm\nu}(x) \\
 \equiv & \displaystyle
(\varSigma u)_{\pm\mu,\pm\nu}(x) \tau^j + \delta_{3j}
(\delta u)_{\pm\mu,\pm\nu}(x) \,,
\end{array}
\ee
where the last equation defines $(\varSigma u)_{\pm\mu,\pm\nu}$. We now
define $(\varSigma C)_{\mu\nu}$ analogously to $(\delta C)_{\mu\nu}$
with $(\delta u)_{\pm\mu,\pm\nu}\to(\varSigma u)_{\pm\mu,\pm\nu}$,
so that we obtain the extension of Eq.~\refc{eq:AWI-lattice} in
the form
\be
\begin{array}{rl}
\displaystyle a^4 \sum_\mu \nabla_x^\mu (\widetilde{J}_A)_\mu^j(x) =
 & \displaystyle \text{\refc{eq:AWI-lattice}} - i \csw a^5 \bp(x)
\sigma_{\mu\nu}H_{\mu\nu}\gf\frac{\tau^j}{2} \p(x) \\
 & \displaystyle - \frac{i}{2} \cem a^5 \Big[ \bp(x) 
\sigma_{\mu\nu} (\varSigma C)_{\mu\nu}(x) \gf\frac{\tau^j}{2} \p(x) +
\frac{1}{2}\delta_{3j} \bp(x) \sigma_{\mu\nu} (\delta C)_{\mu\nu}(x)
\gf \p(x) \Big] \,.
\end{array}
\ee

\section{Multiplicative mass renormalization in magnetic fields}
\label{app3}

In this Appendix we demonstrate that the QCD mass renormalization 
constant is independent of the background magnetic field.
On general grounds, it is expected that the ultraviolet divergent 
renormalization constants of the theory 
do not depend on physical parameters like the magnetic field. 
Here we show this using one-loop perturbation theory in continuum
QCD.\footnote{For a similar calculation at nonzero temperature, see
Ref.~\cite{Bandyopadhyay:2017cle}.} The mass renormalization 
constant is obtained from the fermion self-energy,
\be
i\,\widetilde{\Sigma}(k)=\raisebox{-1.7em}{
\begin{tikzpicture}[ thick,font=\small ]
 \draw[black, style=double, double distance=2.pt, thick] (0,0) -- (0.5,0); 
 \draw[black, style=double, double distance=2.pt, thick] (0.5,0) -- node[below]{p} ++(2.5,0);
 \draw[black, style=double, double distance=2.pt, thick] (3,0) -- (3.5,0); 
 \draw[decorate, decoration={coil,amplitude=4pt, segment length=6pt}, draw=black] (.7,0) to [out=90,in=90] node[above]{$k-p$} ++(2.2,0);
\end{tikzpicture}}
= g^2\int \frac{\dd^4 p}{(2\pi)^4} \,\gamma_\mu \,t^a \,\widetilde{S}(p) \,\gamma_\nu \,t^b \,\widetilde{D}^{ab}_{\mu\nu}(k-p)\,,
\label{eq:sigma}
\ee
where the double line represents the free fermion propagator in the background magnetic field 
and the curly line represents the gluon propagator. The $\mathrm{SU}(3)$ generators are 
denoted by $t^a$. 

In the Schwinger proper time representation~\cite{Schwinger:1951nm}, 
the Feynman-gauge 
gluon propagator reads,
\be
i\widetilde{D}^{ab}_{\mu\nu}(k) = -\delta_{\mu\nu} \delta^{ab} \int_0^\infty \dd t \, e^{-k^2t}\,,
\ee
while the $B$-dependent fermion propagator is taken from Ref.~\cite{Shovkovy:2012zn}
(analytically continued to imaginary proper times and Wick rotated to 
Euclidean space-time with Euclidean Dirac matrices),
\be
i\widetilde S (p) = \int_0^\infty \dd s \,e^{-(m^2+p_\parallel^2)s -p_\perp^2/(qB)\, \tanh(qBs)}
\left[ i\slashed{p}+m -(p_x\gamma_y-p_y\gamma_x) \tanh (qBs) \right] \left[ 1-i\gamma_x\gamma_y \tanh(qBs) \right]\,,
\label{eq:Scomplete}
\ee
where $p_\perp^2=p_x^2+p_y^2$ and $p_\parallel^2=p_z^2+p_t^2$ and throughout we will assume 
that $qB>0$. 

In fact, Eq.~(\ref{eq:Scomplete}) is the 
Fourier transform of the translational invariant part of the propagator, 
and excludes the Schwinger phase. 
Inverse Fourier transforming 
Eq.~(\ref{eq:sigma}) and multiplying by the Schwinger phase gives 
the coordinate space representation of both sides~\cite{Shovkovy:2012zn,Miransky:2015ava}. 
Accordingly, the perpendicular $p_\perp$ are not real 
quantum numbers but referred to as pseudomomenta. 

The propagator can be expanded in powers of the magnetic field.
The consecutive orders read
\be
\begin{split}
i\widetilde{S}_0(p) &= \int_0^\infty \dd s \,e^{-(m^2+p^2)s} \left[ i\slashed{p}+m \right]\,, \\
i\widetilde{S}_1(p) &= -i\,qB\, \gamma_x\gamma_y\int_0^\infty \dd s \,s\,e^{-(m^2+p^2)s} \left[ i(p_z\gamma_z+p_t\gamma_t) +m \right]\,, \\
i\widetilde{S}_2(p) &= (qB)^2 \int_0^\infty \dd s \,s^2\,e^{-(m^2+p^2)s} \left[ (i\slashed{p}+m) p_\perp^2 \frac{s}{3} -i(p_x\gamma_x+p_y\gamma_y)  \right]\,,
\end{split}
\label{eq:S012}
\ee
and can be inserted, order by order in the magnetic field,
into $\widetilde\Sigma$. In color space, we only have the trivial factor
$\sum_a t^at^a = C_F\,\mathbf{1}$ with $C_F=4/3$. The Dirac structure can be 
simplified with conventional techniques.

The $B=0$ contribution to the self-energy reads, after integration over $p$,
\be
i\widetilde{\Sigma}_0(k)=\frac{g^2}{8\pi^2} C_F \int_0^\infty \dd s \int_0^\infty \dd t \frac{1}{(s+t)^2}\left[ 2m-i\slashed{k}\frac{t}{s+t} \right] e^{-m^2s -k^2\frac{st}{s+t}}\,.
\ee
This can be analytically integrated in $t$. The result at low $s$ behaves like $1/s$, 
such that an UV cutoff is necessary at the lower limit of integration $1/\Lambda^2<s$,
and the integral diverges as $\log \Lambda$.
To obtain the mass renormalization constant, we need to 
evaluate $\widetilde{\Sigma}$ at the pole 
of $\widetilde{S}_0$. This amounts to the replacement $\slashed{k}\to-im$ 
(compare to Ref.~\cite{peskin1995introduction} and remember that we are working with
Euclidean metric). We arrive at
\be
i\widetilde{\Sigma}_0(\slashed{k}\to-im) = \frac{3g^2}{16\pi^2} C_F\,m \left[ \log\frac{\Lambda^2}{m^2} -\gamma_E +\frac{5}{6} \right]\,,
\label{eq:SigmaB0}
\ee
which is the well-known one-loop mass renormalization constant in QCD. 
Equation~(\ref{eq:SigmaB0}) 
coincides with Schwinger's result for QED (replacing $g\to q$ and $C_F\to1$).

We proceed with the $\mathcal{O}(qB)$-term. After integration over $p$, we get
\be
i\widetilde{\Sigma}_1(k) = \frac{g^2eB}{8\pi^2} C_F \gamma_x \gamma_y (k_z\gamma_z+k_t\gamma_t) \int_0^\infty \dd s \int_0^\infty \dd t \, \frac{st}{(s+t)^3} \,e^{-m^2s-k^2\frac{st}{s+t}}\,,
\ee
which can be integrated directly over $t$. The result behaves as $s^0$ at low $s$, 
and thus involves no ultraviolet divergences. Analytically integrating over $s$ results in
\be
i\widetilde{\Sigma}_1(k) = \frac{g^2eB}{8\pi^2} C_F\, \gamma_x \gamma_y (k_0\gamma_0+k_3\gamma_3) \frac{1}{k^2} \left[ 1+\frac{m^2}{k^2} \log\frac{m^2}{k^2+m^2} \right]\,,
\ee
which is related to the anomalous magnetic moment of the quark~\cite{Schwinger:1951nm}.

Finally, we consider the $\mathcal{O}((qB)^2)$-term. 
Also, here, we are able to perform 
the $p$-integral, followed by the $t$-integral. The result is linear in
$s$ at low $s$ and is thus again ultraviolet finite.
In fact, since higher powers of the magnetic field always appear with 
higher powers of $s$ [see Eq.~\refc{eq:S012}], 
the ultraviolet region $s\approx0$ is getting 
more and more suppressed as we consider higher and higher 
orders in the magnetic field. 
Altogether, this shows that the $B$-dependent part of the quark self-energy 
is, to all orders in $B$, finite. Therefore, the mass renormalization constant
contains no $B$-dependent divergences and one 
is free to use a $B$-independent renormalization scheme to define
the renormalized mass.

\end{document}